%% file: main.tex
\UseRawInputEncoding
\documentclass[preprint,review,12pt]{elsarticle}
\usepackage{lineno,hyperref}
\usepackage{mathrsfs}
\usepackage{braket}
\usepackage{amssymb,amsmath}

\usepackage{color,xcolor}

\usepackage{amssymb}
\journal{Journal of Information Fusion}

\begin{document}

\begin{frontmatter}

\title{Quantum-inspired Multimodal Fusion for Video Sentiment Analysis}

\author[unipd]{Qiuchi Li\corref{cor1}}
\address[unipd]{University of Padua}
\ead{qiuchili@dei.unipd.it}

% \author[unipd]{Qiuchi Li}
% %\address{University of Padua}
% %\ead{qiuchili@dei.unipd.it}

\author[ou]{Dimitris Gkoumas}
\address[ou]{The Open University}
\ead{dimitris.gkoumas@open.ac.uk}

\author[dku]{Christina Lioma}
\address[dku]{University of Copenhagen}
\ead{c.lioma@di.ku.dk}

\author[unipd]{Massimo Melucci}
% %\address{University of Padua}
\ead{melo@dei.unipd.it}

\cortext[cor1]{I am corresponding author}

% \address[unipd]{University of Padua}
% \ead[unipd]{}

% \author{Anonymous}
% \tnotetext[label0]{This is only an example}

% \author[label1,label2]{Author One\corref{cor1}\fnref{label3}}

% \address[label2]{Address Two\fnref{label4}}

% \cortext[cor1]{I am corresponding author}
% \fntext[label3]{I also want to inform about\ldots}
% \fntext[label4]{Small city}

% \ead{author.one@mail.com}
% \ead[url]{author-one-homepage.com}

% \author[label5]{Author Two}
% \address[label5]{Some University}
% \ead{author.two@mail.com}

% \author[label1,label5]{Author Three}
% \ead{author.three@mail.com}

\begin{abstract}
We tackle the crucial challenge of fusing different modalities of features for multimodal sentiment analysis. Mainly based on neural networks, existing approaches largely model multimodal interactions in an implicit and hard-to-understand manner. We address this limitation with inspirations from quantum theory, which contains principled methods for modeling complicated interactions and correlations. In our quantum-inspired framework, the word interaction within a single modality and the interaction across modalities are formulated with superposition and entanglement respectively at different stages. The complex-valued neural network implementation of the framework achieves comparable results to state-of-the-art systems on two benchmarking video sentiment analysis datasets. In the meantime, we produce the unimodal and bimodal sentiment directly from the model to interpret the entangled decision.
\end{abstract}

% \begin{highlights}
% \end{highlights}


\end{frontmatter}

% \linenumbers


\input{introduction.tex}

\input{background.tex}
\input{related_works.tex}

\input{methodology.tex}
\input{experiment.tex}

\input{results.tex}

\input{conclusion.tex}
\input{acknowledgements.tex}
%% References


\bibliographystyle{elsarticle-num}

\bibliography{mybib}

\end{document}

%% file: introduction.tex
\section{Introduction}
\begin{figure}
\centering
\caption{A motivating example. The sentence ``The pizza is OK'' has different sentiments in different visual-acoustic contexts. One needs to jointly model the signals from three modalities in order to judge the entailed sentiment correctly.}
\includegraphics[width=0.8\textwidth]{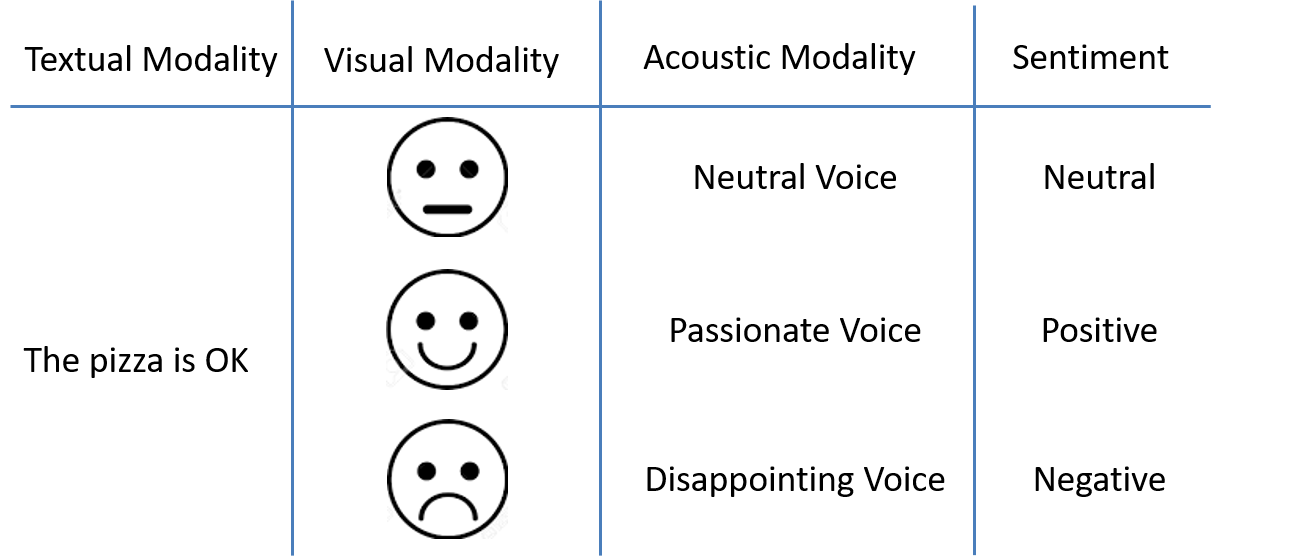}
\label{fig:motivating_example}
\end{figure}

Multimodal sentiment analysis is an emerging topic in natural language processing. It aims at identifying the sentiment of a speaker by gaining clues from multimodal signals, including textual, visual and acoustic channels. From a cognitive point of view, humans express feelings or emotions via different channels. Quite often, a particular sentence entails different sentiments under different visual-acoustic contexts. As shown in Fig.~\ref{fig:motivating_example}, by uttering the sentence ``The pizza is OK'', a speaker could stay unbiased towards the pizza w
dith a blank facial expression and a neutral voice, show his affection to it with a smiling face and a passionate voice, or express a negative attitude with a despising look and a disappointing sound. Consequently, it is not enough to rely solely on textual clues to judge sentiment, and considering visual and acoustic signals is equally essential. Therefore, the core challenge, falls on the fusion strategy that effectively captures the interactions between multimodal signals.

The two main components of multimodal interactions are \textit{intra-modal} and \textit{inter-modal} interactions. Intra-modal interactions refer to the interactions between features within a specific modality, such as the language structure and word dependencies for text or the evolution of human facial expressions over time. Modeling of intra-modal interactions will lead to a unimodal representation and decision on its basis. Inter-modal interactions are the interactions between different modalities on a higher level. The examples above are accounts of a simple inter-modal interaction between visual and acoustic modalities where they are always consistent with each other in terms of sentiment judgment. More often, different modalities contain different sentiment tendencies when judged individually, and a joint decision is in place to fuse the unimodal results.

Prior work mainly models intra- and inter-modal interactions and conduct multimodal fusion on the feature level via neural networks~\cite{zadeh_tensor_2017,liu_efficient_2018,liang_multimodal_2018,zadeh_multi-attention_2018,zadeh_memory_2018,bagher_zadeh_multimodal_2018,pham_found_2019,tsai_multimodal_2019,barezi_modality-based_2019,liang_learning_2019,mai_divide_2019,wang_words_2019,tsai_learning_2019,rahman_integrating_2020}. The neural structures allow the models to learn multimodal interactions typically from a large scale of data in an end-to-end fashion, often leading to satisfactory accuracy. However, the multimodal interactions are implicitly encoded by those models, which adds difficulties to understand the multimodal interactions in human language. Some models do not contain specific components to handle intra-modal and inter-modal interactions explicitly, but jointly model both levels of interactions by fusing multimodal features at each timestamp in a recurrent structure~\cite{zadeh_multi-attention_2018,zadeh_memory_2018,bagher_zadeh_multimodal_2018,liang_multimodal_2018}. Others rely on sequence-to-sequence structure to directly obtain joint representations of a sentence based on its word-level unimodal features~\cite{pham_found_2019,tsai_multimodal_2019}. The tensor-based approaches have also been employed as a mathematical structure for fusing unimodal features either on the word level or on the sentence level~\cite{liu_efficient_2018,mai_divide_2019,zadeh_tensor_2017,liang_learning_2019}.

In all three directions of performing multimodal fusion, the way the modalities interact is often vague and implicit for both levels of interactions~\cite{baltrusaitis_multimodal_2017}. It is closely related to the interpretability issue and the broader concept of explainable AI. Interpretability has become a significant concern for machine learning models. As those models have brought about remarkable performance boosts, researchers are looking for ways to understand the model, in order to know whether we can trust it and deploy it in real work~\cite{lipton_mythos_2018}, or whether it contains privacy or security issues~\cite{holzinger_current_2018}. Existing models in multimodal sentiment analysis heavily rely on neural structures to fuse multimodal data, which often behave like black-boxes with few numerical constraints and purely data-driven assignment. As a result, these models invariably suffer from low interpretability. 

In this paper, we investigate a quantum-inspired approach for fusing multimodal data, in an attempt to provide a principled view of multimodal fusion from a quantum perspective. The inspiration stems from the manifestation of non-classical phenomena in human cognition and decision, which violates classical probability theory but adopts a compact explanation via quantum theory (QT)~\cite{busemeyer_quantum_2012}.  QT has stimulated the successful construction of quantum-inspired models for human cognition-related tasks, such as information retrieval (IR)~\cite{sordoni_modeling_2013,li_quantum-inspired_2018} and language understanding~\cite{wang_semantic_2019,li_cnm:_2019}.  As a typical human cognitive task, however, multimodal sentiment analysis has received little attention from a quantum-inspired viewpoint~\cite{zhang_quantum-inspired_2018,gkoumas_investigating_2018}, due to the challenge in modeling complicated interactions across different modalities in a quantum manner.

We present a novel quantum-theoretic multimodal fusion framework. The framework aims to recognize the sentiment of a multimodal sentence, which consists of word-aligned features of visual, acoustic and textual modalities. Because of the word-aligned format of data, the intra-modal interactions mainly consist of relations between different words in different modalities, while the inter-modal interactions are the interactions between visual, acoustic and textual modalities. Our framework answers the following questions:

\begin{enumerate}
    \item How can we model the intra-modal interactions between different words?
    \item How can we model the inter-modal interactions across different modalities?
    \item How can we make sentiment predictions based on the sentence representation?
\end{enumerate}

The interactions between different words are captured by \textit{quantum superposition} on the feature level. Quantum superposition has been successfully applied to IR~\cite{sordoni_modeling_2013} and natural language processing (NLP)~\cite{wang_semantic_2019,li_cnm:_2019}, with advantages in modeling word dependency and formulating semantic composition. We seek to extend the application  scope from text analysis to a multimodal context, and the quantum superposition of words is instrumented based on the multimodal word representation. 

As for inter-modal interactions, \textit{quantum entanglement} is adopted to fuse unimodal sentiment decisions. Quantum entanglement means that a state in a composite system is so correlated that it cannot be decomposed into products of subsystem states. In analogy to quantum entanglement, for multimodal sentiment analysis, the overall sentiment is perceived through a complicated understanding of textual, visual and acoustic signals that can hardly be viewed as a simple combination of unimodal clues. On the implementation level, this gives rise to explicit tensor product of unimodal sentiment representations. 

In our model, sentiment decisions are made via the concept of~\textit{quantum measurement}, which is a natural choice given the quantum state representation of multimodal sentences. Concretely, an \textit{observable} is introduced to measure the probabilities of the multimodal sentences in the states of main sentiment-related aspects. The probability values are then passed to a fully connected layer to predict the final sentiment. Each sentiment-related aspect is in effect an \textit{entangled state}, indicating that the model views the sentiment judgment as a complicated combination of unimodal decisions, which cannot be simply decomposed into judgments based on each of the three modalities.

We evaluate our model by comparing it against state-of-the-art (SOTA) baselines on two benchmarking multimodal sentiment analysis datasets, namely CMU-MOSI~\cite{zadeh_mosi:_2016} and CMU-MOSEI~\cite{bagher_zadeh_multimodal_2018}. In a fair and comprehensive comparison, our model achieves comparable accuracy to the SOTA models. Furthermore, an illustration of model interpretability is given by elaborating on explaining the quantum process from a classical point of view. In  particular, we present the unimodal and bimodal sentiment judgments implicitly entailed in our model's entangled sentiment predictions.

%% file: background.tex
\section{Preliminaries on Quantum Theory}

Quantum Theory (QT) provides a mathematical interpretation of the microscopic world such as electrons and photons, by formulating the events of this world as subspaces in a vector space with projective geometry. Here we briefly introduce the core concepts of the quantum theory to set the basis of our proposed multimodal fusion framework.

The mathematical formulation of superposition is established on the basis of a \textit{Hilbert Space} $\mathcal{H}$, which is an infinite-dimensional inner product space over the complex field. We adopt the widely-used \textit{Dirac Notations} for a mathematical representation of quantum concepts for consistency with quantum theory.  Essentially,  a complex-valued \textit{unit} vector $\vec{\mu}$ and its conjugate transpose $\vec{\mu}^H$ are denoted as a \textit{ket} $\ket{u}$ and a \textit{bra} $\bra{u}$ respectively. The inner product of two unit vectors $\vert u \rangle$ and $\vert v \rangle$ is a \textit{braket} $\braket{u|v}$, and $\ket{u}\bra{v}$ refers to the complex-valued matrix representing their outer product.

\subsection{Superposition}
~\label{sec:superposition}
As a fundamental concept in Quantum Theory, \textit{superposition} describes the inherent uncertainty in the state of a microscopic particle, such as a photon or an electron. In the microscopic world, a physical property of a single particle, such as location or the momentum, can simultaneously take different values . In the quantum language, the particle is in a \textit{superposition} of multiple mutually exclusive basis states $\{\ket{e_i}\}$, which is a set of complete and mutually orthogonal basis vectors over the complex field. In a two-dimensional Hilbert Space $\mathcal{H}_2$, the basis vectors are denoted as $\ket{0}$ and $\ket{1}$. A general \textit{pure state} is mathematically a linear combination of basis vectors with complex-valued weights such that 
\begin{equation}   
\label{eq:superpostion}
\ket{\phi} = \alpha_0 \ket{0} + \alpha_1 \ket{1}, 
\end{equation} 
where $\alpha_0$ and $\alpha_1$ are complex scalars with $|\alpha_0|^2 + |\alpha_1|^2 = 1$, and $|\cdot|$ is the modulus of a complex number. It follows that $\ket{\phi}$ is a 2-dim unit vector defined over the complex field. When $\alpha_0$ or $\alpha_1$ is zero, then $\ket{\phi} = \ket{0}$ or $\ket{1}$ is a basis state. When $\alpha_0$ and $\alpha_1$ are non-zero complex values, the state $\ket{\phi}$ is said to be a superposition of the states $\ket{0}$ and $\ket{1}$, and the scalars $\alpha_0$ and $\alpha_1$ are amplitudes of the superposition whose squared moduli correspond to probabilities in the measurement, as introduced in Section~\ref{section:measurement}.

\subsection{Mixture}
\label{sec:mixture}
\textit{Mixture} describes the overall state of a set of pure states in probabilistic distribution, i.e., a \textit{mixed system}. The mathematical representation of the mixed system state is a \textit{density matrix}, which is a positive semi-definite square matrix with a unitary trace. Essentially, for a set of pure states $\{\ket{\phi_i}\}_{i=1}^n$ with probability weights $\{p_i\}_{i=1}^n$, the density matrix $\rho$ for the mixed state is computed by
\begin{equation}   
\label{eq:mixture}
\rho = \sum_i^n p_i\ket{\phi_i}\bra{\phi_i},
\end{equation}
Since $\{p_i\}_{i=1}^n$ are non-negative values that sum up to 1, the complex-valued density matrix $\rho$ produced by Eq.~\ref{eq:mixture} is always positive semi-definite with unit trace, i.e., $\rho = \rho^H, tr(\rho) = 1$. The diagonal elements of $\rho$ are always non-negative real values that sum up to 1, while the off--diagonal entries are generally complex values. 

\textit{Mixture} encodes the uncertainty in the distribution of different particles. Different probabilistic distributions of a set of pure states may give rise to the same (i.e., physically indistinguishable) mixed state. For instance, for the set of states $\{\ket{v_1} = [1,0], \ket{v_2} = [0,1], \ket{v_3} = [\frac{\sqrt 2}{2},-\frac{\sqrt 2}{2}], \ket{v_4} = [\frac{\sqrt 2}{2},\frac{\sqrt 2}{2}]\}$, we have $\frac{1}{2}\ket{v_1}\bra{v_1} + \frac{1}{2}\ket{v_2}\bra{v_2} = \frac{I_2}{2} = \frac{1}{2}\ket{v_3}\bra{v_3} + \frac{1}{2}\ket{v_4}\bra{v_4}$ ($I_2$ is the 2 by 2 identity matrix), so the distributions $\{\frac{1}{2},\frac{1}{2},0,0\}$ and $\{0,0,\frac{1}{2},\frac{1}{2}\}$ result in the same state.

It is also worth noting that a pure state can be converted to a density matrix, i.e., $\rho = \ket{\phi_k}\bra{\phi_k}$ for some $k$. Hence the density matrix $\rho$ can be used to represent both a pure and a mixed state in a single Hilbert Space.

\subsection{Measurement}
\label{section:measurement}

\textit{Measurement} is the process of measuring the physical property of a system, such as the momentum or position of a particle. Measurement is associated with a set of possible values. Prior to the measurement, there is uncertainly in the system in that it takes all possible measurement values simultaneously. After the measurement, the system \textit{collapses} onto precisely one value, and the uncertainly on the system state is hence removed. 

The mathematical concept that controls the measurement is an \textit{observable} $\hat{O}$, which is a self-joint square matrix, i.e., $\hat{O} = \hat{O}^H$. Under the eigenvalue decomposition, the eigenstates $\{\ket{\lambda_i}\}$ of $\hat{O}$ form a complete orthogonal basis of the Hilbert Space $\mathcal{H}$, while the eigenvalues $\{\lambda_i\}$ correspond to the possible values the system can take after measurement. 

Suppose a density matrix $\rho$ represents the system before measurement. The probability $p_i$ that the system takes the value of $\lambda_i$ is then computed by Born's rule~\cite{gleason1957measures} as below:

\begin{equation}
\label{eq:trace_measurement}
p_i = tr(\rho \ket{\lambda_i}\bra{\lambda_i}) = \bra{\lambda_i} \rho \ket{\lambda_i} 
\end{equation} 
\noindent Since $\{\ket{\lambda_i}\bra{\lambda_i}\}$ are a complete orthogonal basis, the resulting probabilities $\{p_i\}$ form a classical probability distribution with $\sum {p_i} = 1$. Therefore, after the measurement, at probability $p_i$, one can observe the system taking value $\lambda_i$, and the system collapses onto state $\ket{\lambda_i}$. 

Since the system collapses onto a certain eigenstate $\ket{\lambda_i}$ after measurement, applying the same observable onto the post-measurement state will always lead to the same state $\ket{\lambda_i}$. However, if the same observable $\hat{O}$ is applied to infinite copies of the same system $\rho$, the observed values will then submit to the classical probability distribution $\{p_i\}$. Moreover, the statistical ensemble of the post-measurement states can be represented as a mixed state

\begin{equation}
\label{eq:post_measurement}
\hat{\rho} = \sum_i p_i \ket{\lambda_i}\bra{\lambda_i}
\end{equation}

\subsection{Composite Quantum System}

\textit{Composite Quantum System} is the system consisting of more than individual quantum systems. Let us consider a two-particle system composed of system $A$ and $B$ with Hilbert Spaces $\mathcal{H}_A$ and $\mathcal{H}_B$ respectively. Then the composite system is defined on the Hilbert Space $\mathcal{H}_{AB} = \mathcal{H}_A \otimes \mathcal{H}_B$, which is a tensor product of the two systems. Essentially, for a set of basis states $\{\ket{e_i}_A\}_{i=1}^{dim_A}$ and $\{\ket{f_j}_B\}_{j=1}^{dim_B}$ for systems $A$ and $B$, $\mathcal{H}_{AB}$ is spanned by the basis states  $\{\ket{e_i}_A \ket{f_j}_B\}_{i=1 j=1}^{dim_A dim_B}$. $\ket{e_i}_A \ket{f_j}_B$ is a simplified notation of the tensor product $\ket{e_i}_A \otimes \ket{f_j}_B$, where the tensor product of two matrices $A = [A_{ij}]$ and $B = [B_{kl}]$ is $A \otimes B = [A_{ij}B]$. Following Sec. 2.1 and 2.2, superposition state $\ket{\psi}$ and mixed state $\rho$ can be defined likewise for composite quantum systems, but the dimension of those states is the multiplication of individual system dimensions, e.g. $dim_A \times dim_B$ for the example above.

\subsubsection{Entanglement}
\label{sec:entanglement}
\textit{Entanglement} describes the state of a composite quantum system in such a way that the measurement outcome on a subsystem does have impact to the outcomes of other subsystems~\cite{melucci_introduction_2015}. For example, the polarization of two photons can be entangled so that the polarization of one photon changes the polarization of the other, even though they are a large distance away from each other~\cite{Zeilinger10}.

Mathematically speaking, entanglement means that the state of a composite quantum system is not separable, i.e., cannot be factored as a product of individual system states.  For a pure state $\ket{\psi} \in \mathcal{H}_{AB}$, $\ket{\psi}$ is \textit{separable} if and only if there exists $\ket{\psi^A} \in \mathcal{H}_A$ and $\ket{\psi^B} \in \mathcal{H}_B$ such that  
\begin{equation}
    \ket{\psi} = \ket{\psi^A} \ket{\psi^B},
\end{equation}
otherwise $\ket{\psi}$ is called an \textit{entangled state}. The concept of entanglement also holds for mixed states. A mixed state $\rho \in \mathcal{H}_{AB}$ is separable if and only if it can be written as

\begin{equation}
    \rho = \sum_i w_i \rho_i^A \otimes \rho_i^B
\end{equation}

\noindent where $w_i \in \mathcal{R}$, $\rho_i^A \in \mathcal{H}_A$ and $\rho_i^B \in \mathcal{H}_B$ for any $i$. When $\rho$ cannot be decomposed as above, it is entangled and hence called an \textit{entangled mixed state}.

Entanglement induces a profound difference between quantum correlation, which is computed for one single system, and classical correlation, which is computed for an ensemble. Consider one system in superposed state and one variable.  In QT, the uncertainty of the measurement of the variable is derived from superposition. If the superposed system consists of two subsystems tensored together, the correlation between the observation of the variable in one subsystem and the observation of the variable in the other subsystem is caused by the fact that the two systems are combined in one single system.  When entanglement exists, the (quantum) correlation can violate some statistical inequalities that cannot be violated when classical correlation is computed.

\subsubsection{Reduced Density Matrix}
\label{sec:rdm}
\textit{Reduced Density Matrix} is used to construct representations of subsystems from a composite quantum system~\cite{nielsen_quantum_2011}. Suppose we have a state $\rho \in \mathcal{H}_{AB}$ for a bi-particle system composed of systems $A$ and $B$. We would like to know the statistical equivalence of $\rho$ on only one system $A$, i.e., $\rho^A$. That is to say, we need to find the density matrix $\rho^A$ of system $A$ so that applying any measurement $M \in \mathcal{H}_A$ onto it yields the same result as applying the measurement $M \otimes I$ on the composite system state $\rho$ with system $B$ unchanged. Following Eq.~\ref{eq:trace_measurement}, we have

\begin{equation}
    tr(M \rho^A ) = tr( (M \otimes I_B)\rho), \forall M \in \mathcal{H}_A
\end{equation}

The solution $\rho_A$ of the above equation is obtained by takingthe \textit{partial trace} of $\rho$ over system $B$:

\begin{equation}
   \rho^A = tr_B(\rho)
\end{equation}

\begin{figure}[h!]
  \centering
  \includegraphics[width=0.8\linewidth]{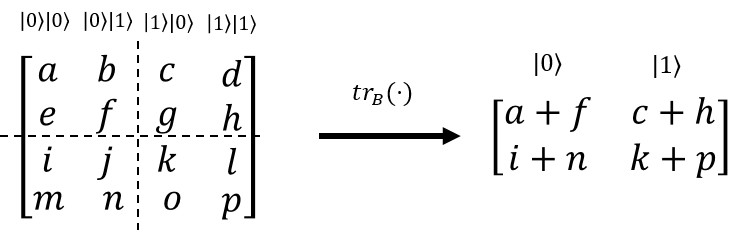}
  \caption{Illustration of partial trace. The left-hand side is a density matrix of a two-qubit system. The partial trace is performed over system $B$. The right hand side shows the resulting 2 by 2 reduced density matrix of system $A$.}
  \label{fig:partial_trace}
\end{figure}

\noindent Where the partial trace is defined as follows: suppose $\rho$ can be expressed as $\rho = \sum_{ijkl} c_{ijkl} \ket{e_i}_A\bra{e_j} \otimes \ket{f_k}_B\bra{f_l}$, then $\rho_A = tr_B(\rho) =  \sum_{ijkl} c_{ijkl} \braket{f_l|f_k}\ket{e_i}\bra{e_j}$. In the matrix form, this stands for computing traces for all blocks corresponding to each subsystem division.  Fig.~\ref{fig:partial_trace} shows the case of taking the partial trace over a two-qubit system state. Since the partial trace operation obeys commutative law, the reduced density matrix can be properly defined over any subset of a composite system of an arbitrary scale (i.e., number of systems). 

The reduced density matrix is indispensable in the analysis of composite quantum systems~\cite{nielsen_quantum_2011}. It has a wide range of applications, of which a typical example is \textit{quantum teleportation} in the quantum communication area. In this work, the reduced density matrix allows us to understand the way unimodal judgments compose the final sentiment judgment. Concretely, by means of a reduced density matrix, we are able to generate the sentiment prediction results for any bimodal and unimodal features, as outlined in Section~\ref{sec:interpretation}.

%% file: related_works.tex
\section{Related Works}

\subsection{Multimodal Sentiment Analysis}

Prior works conduct multimodal sentiment analysis by fusing different modalities either on the feature level or on the decision level.

Decision-level approaches conduct sentiment analysis based on each individual modality first, and then propose different mechanisms to merge unimodal sentiment decisions to the final decision, including averaging~\cite{shutova_black_2016}, majority voting~\cite{morvant_majority_2014}, weighted sum~\cite{evangelopoulos_multimodal_2013} or a learnable model~\cite{glodek_multiple_2011}. Decision-level fusion models are often light-weight, flexible, and scale well to the number of modalities. However, the low-level interactions across different modalities are ignored, which adversely affects performance.

Feature-level methods construct a joint representation by considering the interaction of multimodal data, and perform sentiment classification on its basis. Considering the time-series format of the benchmarking multimodal sentiment analysis datasets, researchers have exploited recurrent neural structures like Long-Short Term Memory (LSTM) to perform a feature-level fusion. Beyond simple concatenation of input features (EF-LSTM) per timestamp or fusion of output hidden units of unimodal LSTMs (LF-LSTM), more complicated fusion strategies have been developed to capture inter-modal interactions. Typically, a hybrid memory is constructed from the hidden units of each modality at a previous timestamp and fed as an additional input of the next timestamp~\cite{liang_multimodal_2018,zadeh_multi-attention_2018,zadeh_memory_2018,bagher_zadeh_multimodal_2018}. 

% Particular strategies include recurrent multi-stage fusion (RMFN)~\cite{liang_multimodal_2018}, multiple attention mechanism (MARN)~\cite{zadeh_multi-attention_2018}, and memory fusion approaches (MFN~\cite{zadeh_memory_2018}, Graph-MFN~\cite{bagher_zadeh_multimodal_2018}). 

Inspired by the notable success achieved by the encoder-decoder structures in sequence-to-sequence (seq2seq) learning, attempts have also been made to ``translate'' the representation of a single sentence under one modality to the representation under another, and take the hidden unit as the joint representation of the sentence~\cite{pham_found_2019,tsai_multimodal_2019}. For example, the multimodal cyclic translation network (MCTN)~\cite{pham_found_2019} applies a seq2seq component to obtain a joint representation of two modalities, and feed the joint representation into another seq2seq structure with the third modality to produce the trimodal representation. Multimodal Transformer (MulT)~\cite{tsai_multimodal_2019} takes a different approach by directly computing the inter-modal representations via the encoder of Transformer~\cite{vaswani_attention_2017}, resulting in a simpler but better-performed network. The structure of Transformer encoder has led to the success of pre-trained language models~\cite{devlin_bert:_2019,yang_xlnet_2019} by effectively capturing word interactions. Based on pre-trained language models, an approach to integrating visual and acoustic features into the pre-trained word-level textual features has recently been proposed~\cite{rahman_integrating_2020} to suit the multimodal context. Since the pre-trained language model well-capture word semantics by training on a large corpus, their multimodal adaptations (MAG-XLNet, MAG-Bert) beat all existing models on multimodal sentiment analysis.

Tensor-based approaches have also been employed to fuse multimodal features. Amongst these models, unimodal features have been merged on the sentence level (TFN~\cite{zadeh_tensor_2017}, LMF~\cite{liu_efficient_2018}, MRRF~\cite{barezi_modality-based_2019}), word level (T2FN)~\cite{liang_learning_2019}, or in a hierarchical manner (HFFN)~\cite{mai_divide_2019} to form a tensorized representation for a multimodal sentence. Based on the representation, fully connected networks~\cite{zadeh_tensor_2017} or tensor decomposition strategies~\cite{liu_efficient_2018,barezi_modality-based_2019} have been employed to generate sentiment decisions.

For the other approaches, Tsai et al.~\cite{tsai_learning_2019} build multimodal representations by factorizing the joint distribution of multimodal data into discriminative factors and modality-specific factors. The constructed Multimodal Factorizational Model (MFM) is essentially an autoencoder, which is capable of integrating any existing algorithm as an encoder that performs multimodal discriminative learning. The additional modality-specific factors bring about performance gain over existing models on six multimodal sentiment analysis datasets. Chaturvedi et al.~\cite{chaturvedi_fuzzy_2019} recently present a new view of the problem, where fuzzy logic is employed to characterize the partial or mixed sentiment. Inspired by this, a convolutional fuzzy sentiment framework is built to map a multimodal sample to fuzzy memberships over four sentiment dimensions and produce sentiment decisions accordingly.

In summary, prior approaches are focused on fusing multimodal information on the feature level. LSTM-based methods do not have an explicit and separate component for handling intra- and inter-modal interactions but combine unimodal features in a per-timestamp manner. Seq2seq-based models construct inter-modal interactions based on correlations of individual word representations in different modalities, largely ignoring the word order information. Tensor-based methods mostly compute the tensor product of unimodal sentence representation to obtain the multimodal sentence representation, while the fine-grained word-level interaction across the modalities is absent. The three types of models suffer from a common interpretability issue, in that the implementation consists of plain neural network components which are hard to understand by humans. The inter-modal hybrid memory, the inter-modal interaction strategy, and the tensor decomposition strategies are crucial to model inter-modal interactions, but they are encapsulated in black-box-like structures without mappings to concrete concepts beforehand.

In contrast, we present a fundamentally novel framework for this particular task. We adopt quantum notions to explicitly tackle intra-modal interactions between words on the feature level and inter-modal interactions on the decision level. At the design phase, the network could be understood as a quantum-like multi-system state preparation and measurement process. This process also allows for a classical interpretation in which the decisions made unimodal and bimodal systems can be produced from the trimodal network, and the entangled pattern over the three modalities can be understood. By implementation, our model conducts word-level tensor fusion based on complex-valued multimodal word representations, which could be viewed as an extended version of tensor-based approaches.

It is worth noting that there is also research focusing on the unaligned data and proposing finer-level modeling of sub-word acoustic and visual dynamics (RAVEN)~\cite{wang_words_2019}.  However, since we only target aligned sentences, it is beyond the scope of this paper. We are also aware of sentic blending~\cite{cambria_sentic_2013}, which is a scalable methodology for fusing multiple cognitive and affective recognition modules in a real-time manner. Our model targets a simpler scenario where the multimodal features of a whole sentence are fed into the model to produce the sentiment label. We will adapt our model to support real-time sentiment analysis in the future.

~\subsection{Quantum-inspired Models for Multimodal Tasks}

Up till now, preliminary studies~\cite{zhang_quantum-inspired_2018,gkoumas_investigating_2018,wang_tensor_2010} have been conducted for addressing multimodal tasks with quantum inspiration. 

Wang et al.~\cite{wang_tensor_2010} build a multimodal image retrieval system by viewing visual-textual documents as non-separable composite systems of image and text and formulating queries as measurements that computes the relevance scores of each multimodal document. However, since the correlation between textual and visual systems are computed by simple statistical methods, the system works no better than a simple concatenation of image and text representations. 

Zhang et al.~\cite{zhang_quantum-inspired_2018} predict sentiment for an image based on its visual content and corresponding textual descriptions. A density matrix is constructed for both an image and a sentence, and the matrices are used to predict the sentiment respectively. Then a decision-level fusion is applied inspired by quantum superposition. The model is not an end-to-end quantum-driven pipeline nor a supervised approach, so it is limited in terms of both efficiency and effectiveness.

Gkoumas et al.~\cite{gkoumas_investigating_2018} investigate the non-classical correlations between decisions made by texts and images by examining the violation of the CHSH inequality, but no violation has been observed.

This work hence contributes to the field of quantum-inspired multimodal analysis in both theory and implementation. An end-to-end quantum-inspired framework for tackling a multimodal task is constructed, and a multimodal fusion is conducted down at the word level. Furthermore, this model is the first to introduce complex values to implement the quantum process into the multimodal context on the implementation level.

~\subsection{Quantum-inspired Models for Text Analysis}

The application of quantum theory to IR begins with the book written by Rijsbergen~\cite{rijsbergen_geometry_2004}, which provides a theoretical framework for formulating IR concepts based on the mathematical formalism of quantum physics. Motivated by this pioneering work, some researchers try to capture the inherent ambiguities in the document relevance judgment. Based on the formulation of a document as a superposition of relevance and irrelevance, the interference between documents is captured by investigating the double-slit experiment~\cite{zuccon_using_2010} or two-round measurements~\cite{zhang_quantum_2016}. Other works~\cite{melucci_towards_2008} seek to exploit entanglement for modeling the interaction between user and document. Apart from the empirical research, preliminary investigations have also been carried out on detecting the quantum phenomena in user relevance judgments~\cite{wang_exploration_2016}.

Other researchers attempt to borrow the quantum probabilistic framework to represent documents and queries. The Quantum Language Model (QLM)~\cite{sordoni_modeling_2013_2} is one of the most successful applications of quantum formalism to IR. QLM builds a density matrix for representing both a document and a query, which is estimated from the pure states of word occurrences and word dependencies in a maximum likelihood estimation algorithm. The documents are ranked according to the VN-divergence of the document density matrices over the query density matrix. QLM achieves an exciting success on ad-hoc retrieval tasks, and spurs later work to extend the original QLM for enhanced performance and expanded applications, such as investigation of quantum entanglement~\cite{xie_modeling_2015} and adoption to session search~\cite{li_modeling_2015}.

Researchers have also applied quantum-inspired models to NLP. A large body of work contributes to quantum-theoretical modeling for compositional meaning~\cite{balkir_distributional_2016,blacoe_quantum-theoretic_2013,coecke_mathematical_2010}. The meaning of each word is represented as a lexical density matrix constructed by combining all the related words in certain contexts. Based on word representations, word ambiguity and semantic entailment have been addressed with a quantum theoretical approach. Zhang et al.~\cite{zhang_quantum_2018} build neural networks to learn high-level interactions of question and answer based on their entangled state representations for question answering (QA).~\cite{li_cnm:_2019,wang_semantic_2019,li_quantum-inspired_2018} leverage quantum \textit{superposition} and \textit{mixture} to model correlations between linguistic features and construct complex-valued language representations by neural networks. The representation leads to improved performance and enhanced interpretability. 

To sum up, quantum-inspired approaches have been widely applied in text modeling, especially in modeling the interactions between fundamental linguistic units, such as words, sentences, concepts, or documents. By employing superposition and entanglement, a principled approach is naturally present to combine the cross-feature interactions with the textual features in a unified representation. However, the prior works are mostly unsupervised models, which do not support learning from a large amount of data. As a result, even though significant improvements have been observed over strong baselines, the models are not as competitive as state-of-the-art neural-based models.

This work follows~\cite{li_cnm:_2019,wang_semantic_2019,li_quantum-inspired_2018} to borrow the concept of superposition and mixture to model intra-modal interactions. The multimodality nature of the data, however, calls for modeling of interactions across different modalities. Therefore, we extend the prior quantum-inspired framework for text to suit the multimodal context and model each modality as a particle in a many-particle system. On its basis, the concept of entanglement is employed as a formulation of inter-modal interactions.

%% file: methodology.tex
\section{Problem Formulation and Notation}

Multimodal sentiment analysis aims to predict the sentiment of a multimodal input sentence. A multimodal sentiment analysis dataset contains $N$ video segments $\mathscr{X}= (X_1,...,X_N)$. Each segment $X_i$ is associated with textual, visual, and acoustic features $X_i = (X_i^t, X_i^v,X_i^a)$ as well as a sentiment label $y_i$.  A common preprocessing step is to align the three modalities of data in terms of words, and then zero-pad the segments to obtain time-series data of the same length $L$. After this step, the textual, visual, and acoustic features of the $i$-th segment are $X_i^t = (t_i^1,...,t_i^L)$, $X_i^a = (a_i^1,...,a_i^L)$ and $X_i^v = (v_i^1,...,v_i^L)$ respectively. Essentially, the multimodal sentiment analysis task is to establish a mapping $f$ that maps each segment $X_i$ to the sentiment $y_i$ entailed in it.

\section{Quantum-inspired Multimodal Fusion Framework}
We now present the quantum-inspired framework for multimodal sentiment analysis. Since Hilbert Space is the mathematical foundation of any quantum-theoretical framework, it is necessary to define the Hilbert Space. In the remaining part of the section, we define the Hilbert Space grounding the proposed framework and introduce the formulation of words, sentences, and sentiment decisions.

\subsection{Multimodal Hilbert Space}

We generally view a multimodal sentence as a composite quantum system of individual modalities. Hence, in our framework, the Hilbert Space is a composition of unimodal Hilbert Spaces for single modalities, hence referred to as \textit{Multimodal Hilbert Space} $\mathcal{H}_{mm}$. In multimodal sentiment analysis, we focus exclusively on the textual, visual, and acoustic modalities. However, it is worth noting that our framework is general and could be adapted to any number of modalities.

\begin{figure}[t]
\centering
\caption{The Multimodal Hilbert Space $\mathcal{H}_{mm}$ composed of textual, visual and acoustic Hilbert Space $\mathcal{H}_t,\mathcal{H}_v,\mathcal{H}_a$. $\ket{e_j^t}, \ket{e_j^v}, \ket{e_j^a}, \ket{e_j^{mm}}$ denotes a basis state of $\mathcal{H}_t,\mathcal{H}_v,\mathcal{H}_a,\mathcal{H}_{mm}$ respectively. }
\label{fig:mhs}
\includegraphics[width=0.7\textwidth]{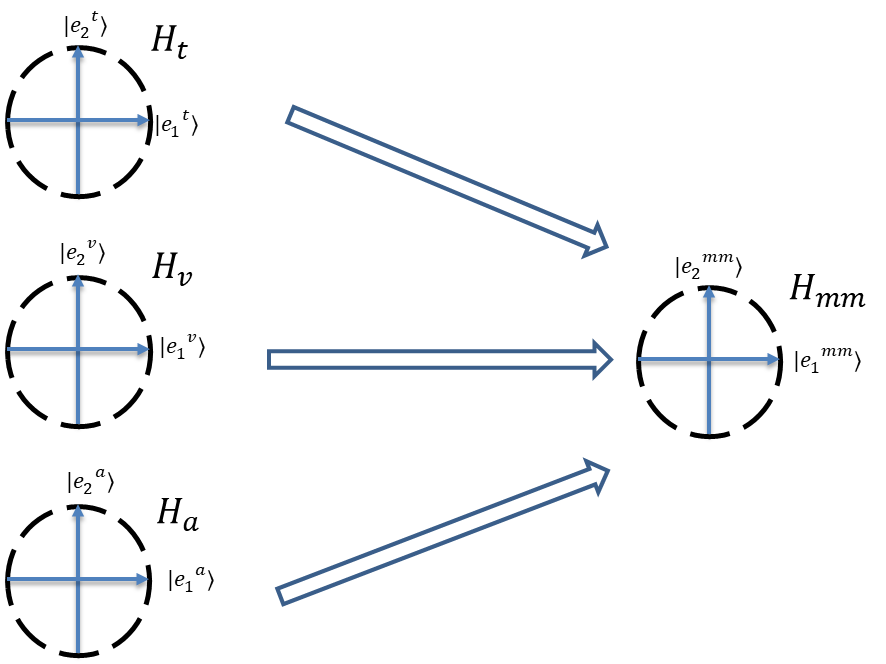}
\end{figure}

Suppose $\mathcal{H}_t,\mathcal{H}_v,\mathcal{H}_a$ denote the Hilbert Space for textual, visual, and acoustic modalities spanned by the basis states $\{\ket{e_i^t}\}_{i=1}^{t_{dim}}$,  $\{\ket{e_j^v}\}_{j=1}^{v_{dim}}$ and $\{\ket{e_k^a}\}_{k=1}^{a_{dim}}$, respectively. $\mathcal{H}_{mm}$ is then expressed as $\mathcal{H}_{mm} = \mathcal{H}_t \otimes \mathcal{H}_v \otimes \mathcal{H}_t$ with a set of basis states $\{\ket{e_i^t} \otimes \ket{e_j^v} \otimes \ket{e_k^a} \}_{i=1 j=1 k=1}^{t_{dim} v_{dim} a_{dim}}$. The basis can be re-written as $\{\ket{e_l^{mm}} \}_{l=1}^{t_{dim}\times v_{dim} \times a_{dim}}$ for simplification purposes, where each $\ket{e_l^{mm}}$ is a tensor product of $\ket{e_i^t}$, $\ket{e_j^v}$, $\ket{e_k^a}$ for some $i,j,k$. Fig.~\ref{fig:mhs} shows the multimodal Hilbert Space in the composition of three individual Hilbert Spaces. 

Classically, each unimodal basis state corresponds to a particular unimodal feature dimension. Hence, a multimodal basis state uniquely reflects a set of unimodal feature dimensions in combination. All possible combinations of unimodal features are taken into consideration by basis states. A sufficient interaction between features from different modalities is therefore in place.

\subsection{Word State}
\begin{figure}[t]
\centering
\caption{Multimodal Word Representation. Each color indicates one word. The multimodal word state $\ket{w}$ for word $w$ is a tensor product of its unimodal states $\ket{w^t}$, $\ket{w^v}$ and $\ket{w^a}$.}
\label{fig:mwr}
\includegraphics[width=0.8\textwidth]{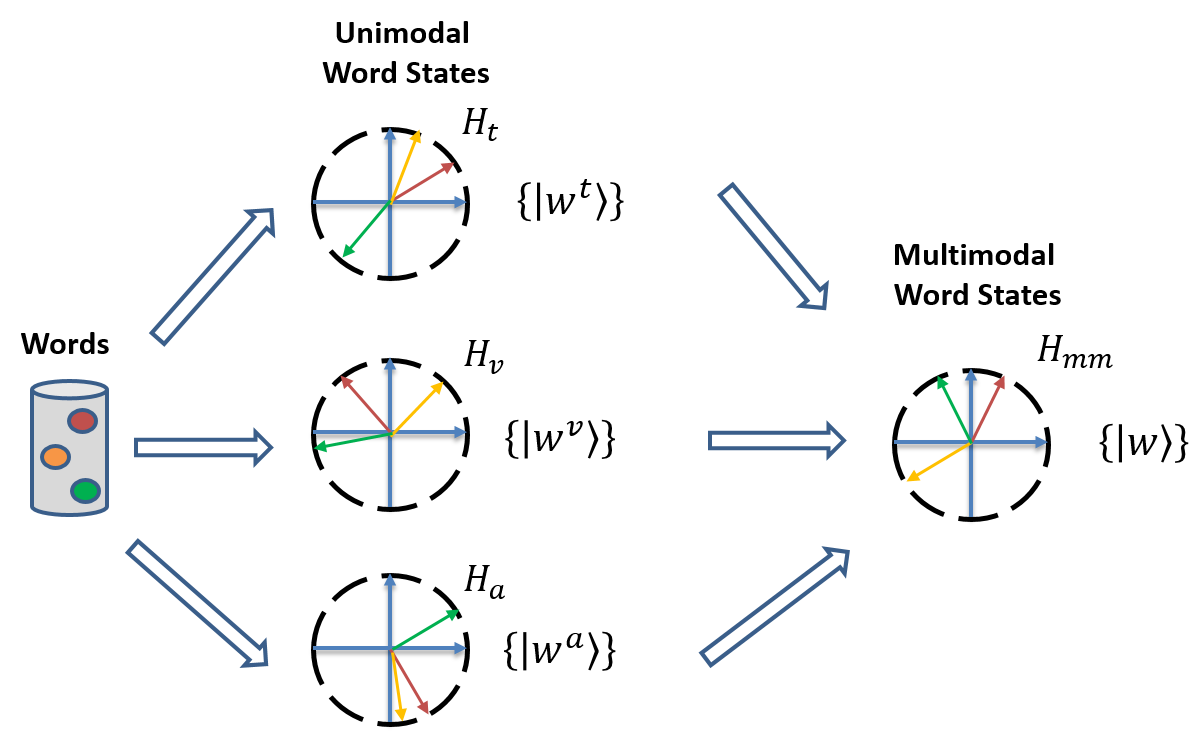}
\end{figure}

A word $w$ is formulated as a pure state $\ket{w}$ on $H_{mm}$. Since a word is associated with a textual, visual and acoustic feature vector, we are able to construct its unimodal state representation $\ket{w^t}$, $\ket{w^v}$ and $\ket{w^a}$ in $\mathcal{H}_t$, $\mathcal{H}_v$, $\mathcal{H}_a$ respectively. It is then an open issue to construct $\ket{w}$ based on the respective unimodal states of $w$. In this work, we assume $\ket{w}$ to be a product state of unimodal states, i.e., $\ket{w} = \ket{w^t} \otimes \ket{w^v} \otimes \ket{w^a}$, as shown in Fig.~\ref{fig:mwr}. This simple strategy is employed for the reasons below: 

\begin{itemize}
    \item By implementation, it gives rise to a tensor-based fusion of multimodal signals, which is believed to be a meaningful and useful approach to capture inter-modal interactions~\cite{liu_efficient_2018,zadeh_tensor_2017,barezi_modality-based_2019}. In particular, it explicitly aggregates features of three modalities by means of multiplication, while other models instead rely on additional structures to fuse unimodal features in a more implicit manner.
    
    \item When uttering one word or one sentence, a person may aim at expressing different sentiments under different situations. A single word has different multimodal representations under different visual-acoustic contexts based on word-dependent textual representation and word-independent visual and acoustic representations. As a result, different sentiments of a specific word or sentence can be accounted for by this multimodal word representation. 

\end{itemize}

\subsection{Sentence State}

\begin{figure}[t]
\centering
\caption{Our framework. Each colored ball indicates a word in the multimodal sentence, represented as a unit vector of the same color in the Multimodal Hilbert Space. The sentence is represented by a mixed state visualized as a black ellipse. The eigenstates of the observable are unit vectors in black color. The squared length of the intersection between each unit vector and the ellipse (in red) is the measurement probability for the respective eigenstate. The sentence sentiment representation is composed of all probability values represented by red balls.}
\includegraphics[width=\textwidth]{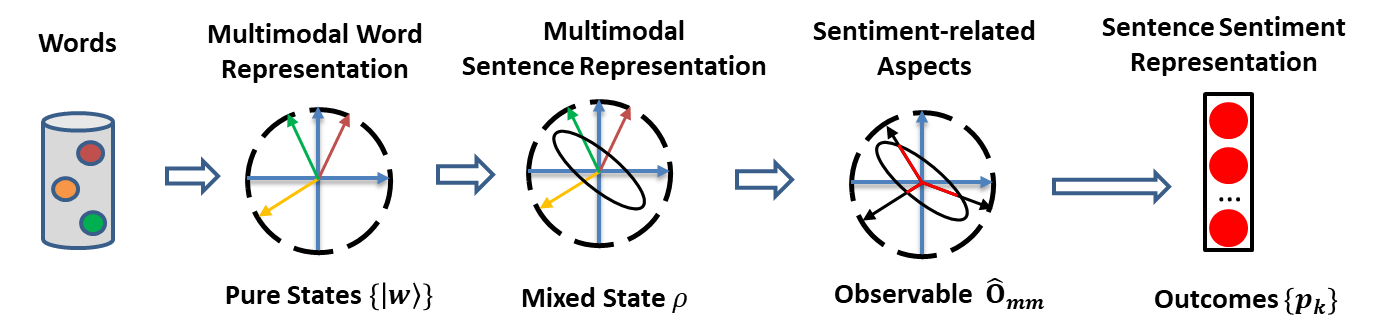}
\label{fig:architecture}
\end{figure}

We formulate a sentence as a mixture of individual word states $\{\ket{w}\}$ in the sentence. The mixed state $\rho \in \mathcal{H}_{mm}$ of a sentence is produced by the individual word states in the sentence in the form of a weighted quadratic summation:

\begin{equation}
    \rho = \sum_i \lambda_i\ket{w_i}\bra{w_i}    
\end{equation}

\noindent where $\{\lambda_i\}$ are convex coefficients, i.e., $\sum_i {\lambda_i} = 1$ in order to guarantee $Tr(\rho)= 1$. $\lambda_i$ is a word-dependent weight that reflects the importance of the word $w_i$ in the sentence. The sentence mixed state $\rho$ is visualized as an ellipse constructed by unit vectors of words in the sentence in Fig.~\ref{fig:architecture}. The ellipse representation is due to the fact that a density matrix assigns a probability measure on the Hilbert Space from the quantum probability point of view. Please refer to~\cite{sordoni_modeling_2013} for a detailed explanation.

Even though a density matrix is constructed from a particular set of word weights, it corresponds to many possible mixture weights of the same set of words (Sec.~\ref{sec:mixture}). As a result, it is capable of formulating different word combinations under different contexts. As a probability measure on the Multimodal Hilbert Space, the density matrix is a sentence representation in terms of unimodal features in combination from a classical perspective. The representation is a separable state rather than an entangled state, because $\rho$ can be re-written as

\begin{eqnarray}
\label{eq:separable}
    \rho &=& \sum_i \lambda_i(\ket{w_i^t} \otimes \ket{w_i^v} \otimes \ket{w_i^a})(\bra{w_i^t} \otimes \bra{w_i^v} \otimes \bra{w_i^a})\\ \nonumber
        &=& \sum_i \lambda_i(\ket{w_i^t}\bra{w_i^t}) \otimes (\ket{w_i^v}\bra{w_i^v}) \otimes (\ket{w_i^a}\bra{w_i^a})\\ \nonumber
        &=& \sum_i \lambda_i \rho_i^t \otimes \rho_i^v \otimes \rho_i^a
\end{eqnarray}

\noindent where $\rho_i^m= \ket{w_i^m}\bra{w_i^m} \in \mathcal{H}_m$ for $m \in \{t,v,a\}$. From Sec.~\ref{sec:entanglement}, $\rho$ is separable with respect to the three unimodal Hilbert Spaces by definition. Consequently, the framework considers word-level interactions via the concepts of mixture and superposition on the feature level, while the interactions across different modalities are largely absent from the feature level. Instead, the inter-modal interactions are implemented in the sentiment decision process, as outlined in the next paragraphs.

\subsection{Sentiment Measurement}
Based on the multimodal sentence representation, a component is needed to operationalize the sentiment judgment process. To this aim, we link sentiment judgment to quantum measurement and ``measure'' the ``sentiment state'' of a multimodal sentence.

We hypothesize that there are $K$ sentiment-related aspects or topics, such as the aspects of aspect-based sentiment analysis. The sentence will collapse onto one of them after the measurement. The probabilities over the aspects after repeating the measurements can be seen as a sentimental characterization of a multimodal sentence, which could be used to determine the sentence sentiment.

Mathematically, the multimodal observable $\hat{O}_{mm}$ is associated with a set of aspect ids $\{k\}_{k=1}^K$ as eigenvalues and a set of aspect representations $\{\ket{v_k}\}_{k=1}^K$ as eigenstates. Hence $\hat{O}_{mm}$ can be expressed as 

\begin{equation}
\label{eq:multimodal_observable}
    \hat{O}_{mm} = \sum_k k\ket{v_k}\bra{v_k}.
\end{equation}

After the experiment, the multimodal sentence $\rho$ will collapse onto the $k$-th aspect at a likelihood of $p_k$: 

\begin{equation}
\label{eq:sentiment_measurement}
    p_k = \bra{v_k}\rho\ket{v_k}
\end{equation}

The final sentiment judgment is given based on the clues from each sentiment-related aspect, and the probability values $\{p_k\}_{k=1}^K$ are taken to generate the sentence sentiment. In Fig.~\ref{fig:architecture}, a set of unit-norm vectors is associated with the observable $\hat{O}_{mm}$. The squared lengths of their intersections with the density matrix ellipse $\rho$ are the measurement probabilities $\{p_k\}_{k=1}^K$ that reflect sentence sentiment.

It is worth noting that each sentiment-related aspect is a pure state $\ket{v_k}$, which is always (i.e., at probability 1) an entangled state of the three unimodal systems. Hence, the aspects are abstract concepts over the whole multimodal space that can hardly be mapped to human-understandable notions. Instead, each aspect can be seen as a multimodal sentiment decision \textit{in entanglement of unimodal sentiment decisions}. The observable $\hat{O}_{mm}$ is uniquely represented by the eigenstates $\{\ket{v_k}\}_{k=1}^K$. In the rest of the paper, we use $\{\ket{v_k}\}_{k=1}^K$ and $\hat{O}_{mm}$ interchangeably to represent an observable.

The way unimodal sentiment decisions are aggregated can be displayed with the help of a reduced density matrix. The reduced density matrix allows us to obtain the statistically equivalent observable for bimodal and unimodal systems so that the decisions entailed in the trimodal system can be inferred by applying the observable onto the respective sentence representation. The details of this process are introduced in Sec.~\ref{sec:interpretation}.

\section{Complex-valued Network for Multimodal Sentiment Analysis}

\begin{figure}[t]
\centering
\caption{The quantum-inspired multimodal fusion network. The multimodal word states are obtained via complex-valued multimodal word embedding. The local context states are constructed from individual word states under the global weighting and local mixture strategy. The multimodal observable is applied to each context state in the measurement step, and the obtained probability matrix is row-wise max-pooled and passed to a neural network to produce the final sentiment.}
\label{fig:qmf_network}
\includegraphics[width=\textwidth]{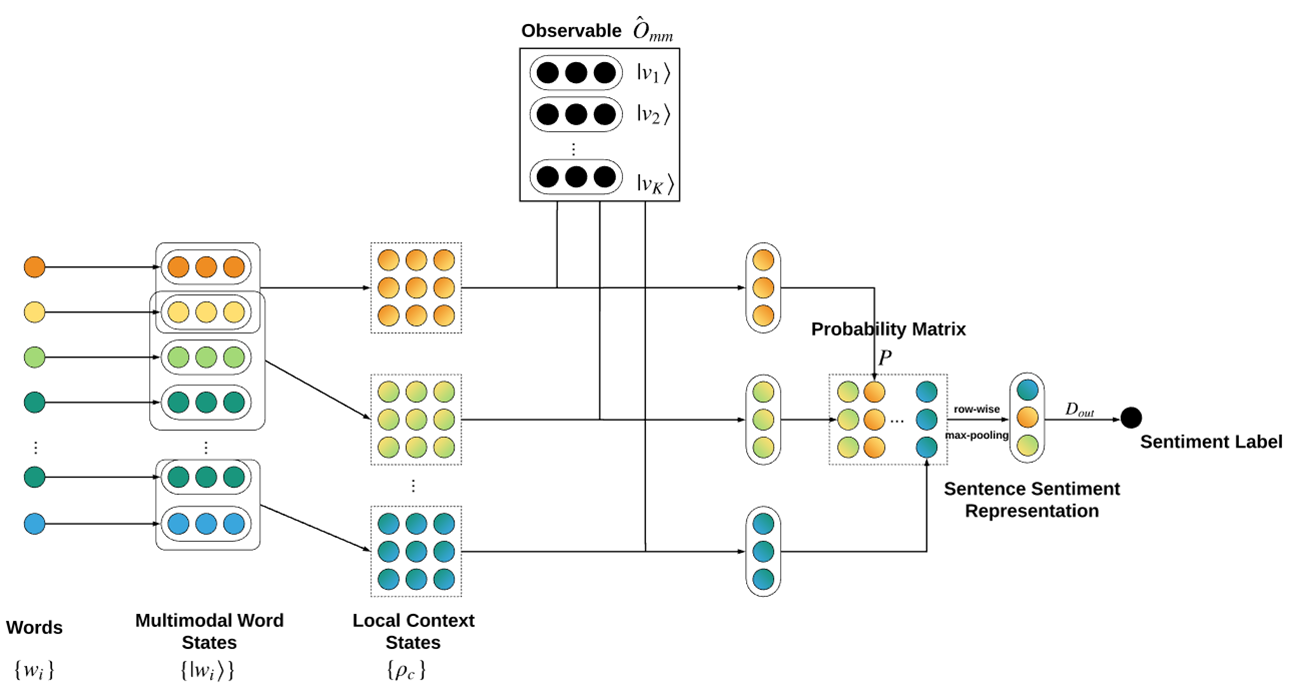}
\end{figure}

This section outlines the neural network implementation of our quantum-inspired multimodal fusion framework for multimodal sentiment analysis. Complex values are pivotal to the formulation of quantum concepts, so our network is composed of complex-valued units as an authentic formulation of the quantum-inspired multimodal fusion process. Fig.~\ref{fig:qmf_network} shows the architecture of the network. Next, we introduce the way to handle complex values for each network component, so that the network weights could be learned in the same way as any classical neural networks.

\subsection{Complex-valued Multimodal Word Embedding}
~\label{sec:embedding}
As previously introduced, the multimodal word state is $\ket{w_i} = \ket{w_i^t} \otimes \ket{w_i^v} \otimes \ket{w_i^a}$ in the Multimodal Hilbert Space. The task is to map real-valued input features to complex-valued unit vectors for each word under each modality. To this aim, we adopt the modulus-argument form for a complex number. Each unimodal state for a word $w$ is represented as  
\begin{eqnarray}
 \ket{w} &=& r_1e^{i\theta_1}\ket{e_1}+...+r_ne^{i\theta_n}\ket{e_n} \\ \nonumber
        &=& [r_1e^{i\theta_1},...,r_ne^{i\theta_n}],
\end{eqnarray}

\noindent where $i$ is the imaginary number satisfying $i^2 = -1$, the moduli $ R = [r_1,...,r_n]$ form a real unit vector, and the arguments $\Theta = \{\theta_1,...,\theta_n\}$ are in $[-\pi,\pi]$ each. In the modulus-argument form, any operation on the complex numbers will lead to a non-linear combination of the constituent moduli and arguments. If the moduli and arguments could be appropriately assigned with different features, a non-linear feature combination is naturally produced.

Different policies are employed to assign the moduli and arguments from the input features for different modalities. The textual modality possesses a word-dependent distributed representation, while the features for non-textual modalities are word-independent and non-trainable. Therefore, the moduli $R^t = [r_1^t,...,r_n^t]$ for $w$ are constructed from the pre-trained word embedding $E(w)$ via a deep neural network $D_t$, while the moduli of visual and acoustic modalities $R^v, R^a$ are obtained via deep neural networks $D_v$ and $D_a$ from the respective input feature vectors $V_w$ and $A_w$ (note that they do not depend on word $w$). Precisely, $R^t = N(D_t(E(w)))$, $R^v = N(D_v(V_w))$, $R^a = N(D_a(A_w))$ with $N(\cdot)$ as the vector $L2$-normalization function. In line with the Tensor Fusion Network~\cite{zadeh_tensor_2017}, $D_t$ is composed of an LSTM layer followed by two fully connected layers, while $D_v$ and $D_a$ are three stacked fully connected layers. They mainly serve as dimension reduction models, ensuring that the dimensionality of the multimodal Hilbert Space is computationally affordable. Moreover, the textual LSTM structure memorizes the sequence information, complementing the quantum-inspired framework that ignores word order. This step produces a low-dimensional representation $R^t$, $R^v$ and $R^a$ for the moduli. They are then unit-normalized to meet the unit-norm constraints. 

The way to initialize arguments for each modality is as follows: for textual modality, we initialize the arguments of sentiment words regarding their respective sentiment polarity. In particular, a positive word is initialized with a zero vector, and a negative word is initialized with a vector of $\pi$, while non-sentiment words are assigned with a vector of $\pi/2$ for their respective textual arguments. The assumption behind this is that the individual word sentiment influences the sentence sentiment, and we aim at leveraging the word sentiment by linking it with the textual arguments. Here we map an argument to the sentiment with the cosine function so that the arguments of $\pi$, $\pi/2$, 0 are mapped to -1,0,1, indicating a negative, neutral and positive sentiment respectively. Since only a rough estimation of word sentiment is present in a sentiment dictionary, it is used as initial values of the arguments subject to fine-tuning together with the other network components. For the non-textual modalities, the arguments $\Theta^v$ and $\Theta^a$ are set to be word-dependent. Even though the non-textual representation is word-independent, different representations of the same word may still share some information that possibly helps to make the sentiment judgment. Hence we build the quantum-inspired framework to learn the arguments $\Theta^v$ and $\Theta^a$ respectively based on unimodal features. The learned arguments are used as initial values of arguments that are fine-tuned on the trimodal data.

\subsection{Mixture}

In the previous section, we have outlined the formulation of a sentence as a mixture of individual words following Eq.~\ref{eq:mixture}. To adapt this step to the multimodal sentiment analysis scenario, one needs to answer the crucial question of determining the word-dependent weights $\{\lambda_i\}$ in Eq.~\ref{eq:mixture}. Furthermore, the sentiment of the sentence is often determined by local contexts, i.e., consecutive words within local windows, rather than the whole sentence. Therefore, another issue falls on the identification of contexts that provide crucial clues to judge the sentiment.

In this paper, we address both issues through \textit{global weighting and local mixture}, which has been taken for constructing text-based language representation~\cite{li_cnm:_2019}. Essentially, we assign a global weight to each word and use the global weights to determine the density matrices for local word contexts. As for the weighting scheme, the weight $\Lambda_i$ of a the word $w_i$ should be composed of its weights under all three modalities. Hence, we apply a weighted sum of unimodal weights to compute $\Lambda_i$:

\begin{equation}
\label{eq:global_weighting}
    \Lambda_i = \beta_t\Lambda_i^t + \beta_v\Lambda_i^v + \beta_a\Lambda_i^a
\end{equation}

\noindent where $\Lambda_i^t = ||D_t(E(w_i))||_2 $, $\Lambda_i^v = ||D_v(V_{w_i})||_2$, $\Lambda_i^a = ||D_a(A_{w_i})||_2$ are the $L2$-norms of the contracted textual, visual and acoustic feature vectors of $w_i$ respectively. $\{\beta_m \in [0,1], m \in \{t,v,a\} \}$ are modality-specific weights that sum up to 1. 

The weighting scheme is followed by the local mixture of words in the multimodal sentence. Specifically, a set of local contexts are identified, and the words in each context $c$ are mixed in a quantum manner (i.e., Eq.~\ref{eq:mixture}) to produce a density matrix $\rho_c$. The mixture weights are produced by softmax-normalizing the word weights within each context so that the outcome of the mixture is always a legal density matrix. The approach to extract local contexts from the sentence is an open issue. In this work, we apply sliding windows of varying lengths through the whole sentence, each producing a density matrix representing a local n-gram. Hence, rather than a single density matrix, a set of matrices are produced by the local mixture component. In the measurement step, the most representative contexts are identified in a data-driven fashion,as outlined in Sec.~\ref{sec:measurement}.

It is worth noting that the L2-norms of feature vectors are used to fit the construction of complex-valued word embedding. In order to ensure that each unimodal representation can be interpreted as a pure state, vector L2-normalization is applied, and vector norms are hence discarded. The vector norm somehow reflects the semantic intensity, which may be indicative of the combination of words in a local context. From a quantum perspective, the absolute number of each pure state should be considered when mixed together.

\subsection{Measurement}
\label{sec:measurement}
The measurement component needs to handle a set of $C$ density matrices $\{\rho_c\}$ for local contexts, and identify the discriminating contexts to sentiment classification. To achieve this purpose, a single observable $\{\ket{v_k}\}_{k=1}^K$ is performed to the set of density matrices, each generating a set of probability values via Eq.~\ref{eq:trace_measurement}. As a result, a $K$-by-$C$ matrix of probability values is produced by the measurement, each entry corresponding to the likelihood of a local context collapsing to an eigenstate. Then a row-wise maximum pooling is conducted to get the most similar local context for each of the K sentiment-related aspects. The respective probabilities, i.e., the K maximum probability values, are treated as the sentence sentiment representation. A neural network $D_out$ is built on its basis to produce the final sentiment prediction of the multimodal sentence.

We aim to learn the eigenstates or sentiment-related aspects $\{\ket{v_k}\}_{k=1}^K$ from the data, as it is difficult to map them to concrete notions beforehand. A deviation from the standard definition of observable is then employed: the set of eigenstates do not necessarily form an orthonormal basis of the Multimodal Hilbert Space, but are instead of a predefined number of $K$ and not hard-coded as orthogonal to each other. The reasons are two-fold. On the one hand, different abstract sentiment-related aspects are not necessarily independent of each other in practice. On the other hand, it is not computationally affordable to ensure mutual orthogonality of measurement states during training, even though there are already algorithms for training mutually orthogonal vectors~\cite{arjovsky_unitary_2016,wisdom_full-capacity_2016}.

\subsection{Network Learning}
The network weights include word embeddings $E$, arguments $\Theta_t$, $\Theta_a$, $\Theta_v$, modality-specific weights $\{\beta_m \in [0,1], m \in \{t,v,a\} \}$, observable $\{\ket{v_k}\}_{k=1}^K$, and neural network weights $D_t, D_a, D_v$ and $D_{out}$. Note that the mixture step does not contain any trainable weights.

As for the initialization,  $E$ is initialized with existing word embedding. The textual arguments $\Theta_t$ is initialized in a word sentiment related manner as introduced in Sec.~\ref{sec:embedding}. In order to initialize the visual and acoustic arguments, we pre-train the framework on respective unimodal data consisting of a dimension reduction network, a global mixture of all words in the sentence, a measurement component and the output network $D_{out}$ with random initialized argument. The eigenstates in the observable have random-initialized arguments and random-initialized unit-norm moduli. 

During training, the moduli and arguments of complex-valued inputs are trained separately with unit-norm constraints imposed on the moduli part. The intermediate complex-valued layers are implemented with real and imaginary parts for inputs and outputs, in order to back-propagate the loss function to real and imaginary parts separately. 

\subsection{Network Interpretation}
\label{sec:interpretation}

Our network captures multimodal interactions by borrowing concepts from quantum theory. For the quantum-like process to be understandable for human beings, we propose an approach to interpret the network.  Essentially, the model captures word interactions via superposition and intermodal interactions through entanglement. Both levels of interactions could be explicitly understood from the learned trimodal model as follows:

I) The unimodal and bimodal decisions entailed in the learned model can be computed for a target sample. The unimodal word states and weights can be computed from the learned model, allowing us to compute global word weights via Eq.~\ref{eq:global_weighting} for any subset of the three modalities, and then the mixed state of any local context on its basis. The corresponding observable for the respective modalities is computed by taking the reduced density matrix (Sec.~\ref{sec:rdm}) of the learned eigenstates so that the measurement can be applied to the set of obtained density matrices. The probabilities are row-wise max-pooled and passed to the learned $D_{out}$ to generate the sentiment label for the target subset of modalities.

II) The multimodal sentiment judgment for any word or word combination can be inferred from the learned model. With the learned observable and output network $D_{out}$, the sentiment label for any density matrix $\rho \in \mathcal{H}_{mm}$ can be produced. A word adopts a density matrix representation (Sec.~\ref{sec:mixture}). The density matrix of any combination of words, such as local contexts, can also be computed as a mixture of word states, with mixture weights being softmax-normalized global word weights. Hence we can check the sentiment for each word or word combination determined by the learned model.

The point that the learned model could be directly leveraged to generate results for part of the data is crucial to address the interpretability issue because that refers to the model's authentic behavior. When the models require re-training on the subset of data, on the other hand, the result cannot be safely interpreted as the performance of the original model anymore. In the multimodal sentiment analysis context, if a trimodal network needs to be re-trained to predict sentiments for unimodal or bimodal data, it will remain doubtful whether the results could be used to ``interpret'' the behavior of the original trimodal model on the unimodal or bimodal systems.

However, instead of directly taking the learned network to give predictions, most prior work in this field requires re-training the model based on unimodal and bimodal data. In particular, LSTM-based approaches involve concatenations of unimodal hidden units to produce the inter-modal dynamics, so one cannot directly apply the component to a bimodal and unimodal case due to the dimension inconsistency. In MulT and tensor-based approaches, the shape of multimodal sentence representation is relevant to the number of modalities, so the neural structures should be re-trained to predict sentiments based on bimodal and unimodal representations. MCTN has a single-directional structure for the second seq-to-seq component, so it could only be used to predict part of the bimodal and trimodal sentiment, depending on the order of the modalities put into modeling. 

To the best of the authors' knowledge, LMF is the only prior model that facilitates direct computation of unimodal and bimodal sentiments from the learned trimodal network. Unfortunately, an analysis of such a property is missing from the original LMF paper. This work identifies this property of our model, and presents the exact prediction results of the trimodal network on unimodal and bimodal data in Sec.~\ref{sec:inter_analysis}.

%% file: experiment.tex
\section{Experimental Setup}

The experiments are conducted on two state-of-the-art benchmarking video sentiment analysis datasets, namely CMU-MOSI~\cite{zadeh_mosi:_2016} and CMU-MOSEI~\cite{bagher_zadeh_multimodal_2018}.  CMU-MOSI is a human multimodal sentiment analysis dataset consisting of 2,199 short monologue video clips (each lasting the duration of a sentence). Acoustic and visual features of CMU-MOSI are extracted at a sampling rate of 12.5 and 15 Hz respectively, while textual data are segmented per word. CMU-MOSEI is a sentiment and emotion analysis dataset made up of 22777 movie review video clips taken from YouTube. The unaligned CMU-MOSEI sequences are extracted at a sampling rate of 20 Hz for acoustic and 15 Hz for visual signals. The visual and acoustic modalities are then aligned with the words~\cite{yuan_speaker_2008}. Human annotators are recruited to label each sample with a sentiment score from -3 (strongly negative) to 3 (strongly positive) for both datasets. The splits of the datasets are given in Tab.~\ref{table:datasets}. 

\begin{table}[t]
\begin{center}

\resizebox{0.5\textwidth}{!}{
\begin{tabular}{l|l|l|l}%{|p{1.5cm}<{\centering}|p{1.5cm}<{\centering}|p{1.8cm}<{\centering}|r|}
 \hline\hline
%  $\Delta$
Dataset & Train & Test & Validation \\ \hline

CMU-MOSI & 1284 & 686  & 229 \\ \hline
CMU-MOSEI & 16265 & 4643 & 1869 \\ \hline\hline
\end{tabular}
}

\caption{Splits for CMU-MOSI and CMU-MOSEI datasets.}
\label{table:datasets}

\end{center}
\end{table}

We notice that different versions of the same dataset are used in the literature, containing different sequence lengths and feature dimensions. For a transparent and reproducible experiment, we clarify our approach to clean and merge the datasets. 

We begin with the datasets in the Github repository of the MulT~\cite{tsai_multimodal_2019} model, which only contains word vectors instead of original words. Since our model requires word-specific embedding and weighting strategies, we decipher words from the given word vectors by cross-checking with different versions of pre-trained 300-dimensional glove embeddings. It turns out that CMU-MOSI and CMU-MOSEI use different dictionaries for the glove embeddings, which leads to unidentified words in CMU-MOSEI. In order to acquire the missing words, we refer to the version of CMU-MOSEI data from CMU-Multimodal SDK~\footnote{https://github.com/A2Zadeh/CMU-MultimodalSDK.git}, which contains the original words but some important words are missing for many samples. By merging the two versions of data, we are able to obtain a clean and complete version of CMU-MOSEI with original words as the textual features. The final datasets have a 35-dim feature and a 74-dim acoustic feature for each word. The lengths of sentences are normalized to 50. The datasets are available upon request.

To evaluate the proposed model, we conduct a comprehensive comparison with the following baseline models:

\begin{enumerate}
    \item \textbf{Early-Fusion LSTM (EF-LSTM)}. This model concatenates the input textual, acoustic and visual features at each timestamp, and builds an LSTM to construct sentence-level multimodal representation. The last hidden state is taken to predict the sentiment. It is broadly used as a baseline model by prior works.
    
    \item \textbf{Late-Fusion LSTM (LF-LSTM)}. This model builds an LSTM for textual, acoustic and visual inputs separately, and concatenates the last hidden state of the three LSTMs as the sentence-level multimodal representation. It is taken to predict the sentiment. It is broadly used as a baseline model by prior works.
    
    \item \textbf{Multi-Attention Recurrent Network (MARN)}~\cite{zadeh_multi-attention_2018}. This model captures inter-modal dynamics at each timestamp. A multi-attention block is built to construct an inter-modal representation based on hidden states of the previous timestamp and fed into the inputs of the current timestamp. The inter-modal representation and hidden states of the last timestamp are concatenated as the multimodal sentence representation and used to classify the sentiment.
    
    \item \textbf{Memory Fusion Network (MFN)}~\cite{zadeh_memory_2018}. This is a typical memory fusion network that builds a multimodal gated memory component, and the memory cell is updated along with the evolution of the hidden states of three unimodal LSTMs. The final memory cell is concatenated with the last hidden states of unimodal LSTMs as the multimodal sentence representation used to classify the sentiment.
    
    \item \textbf{Tensor Fusion Network (TFN)}~\cite{zadeh_tensor_2017}. This model extracts unimodal sentence representations with different neural network structures, and computes the tensor product of the unimodal vectors as the multimodal sentence representation. An additional value of 1 is appended to each unimodal vector, such that the final tensor product also entails interactions among a subset of the modalities. The tensor product is then flattened and used to classify the sentiment.
    
    \item \textbf{Low-rank Multimodal Fusion (LMF)}~\cite{liu_efficient_2018}. This model adopts the same procedure as TFN to construct the multimodal representation but then computes the inner product between the three-order tensor with a weight tensor with a low separation rank to produce a low-dimension vector output. With tensor decomposition, the inner product is converted to the inner product of unimodal features with respective weights. The vector is used to classify the sentiment.
    
    \item \textbf{Multimodal Transformer (MulT)}~\cite{tsai_multimodal_2019}. This model encodes inter-modal attention into an enriched unimodal representation. The multi-head attention block in~\cite{vaswani_attention_2017} is adopted to learn the inter-modal attention. The unimodal representations are concatenated as the multimodal representation, which is used to predict the sentiment of the sentence.
    
\end{enumerate}

The above baselines have good coverage of the main types of models on the multimodal sentiment analysis task, including the vanilla LSTM early-fusion and late-fusion strategies, existing LSTM-based inter-modal dynamic fusion models, tensor-based approaches and a seq-to-seq-based model. MFM~\cite{tsai_learning_2019} is excluded from the experiment because it is a generic structure, and all the aforementioned models can be integrated as an encoder of the structure. Therefore, it is fair to experiment with all the models in the absence of MFM to check their effectiveness. 

To ensure a fair comparison, we apply uniform experimental settings for different models. For all models but LMF, we build the same neural structure for predicting sentiment based on multimodal sentence representation, which is composed of two fully connected layers with a rectified linear unit (ReLU) as the activation function. Since a low-dimension (around 5) vector representation is present for LMF as the multimodal representation, one fully connected layer is used for sentiment prediction for LMF. $L1$-loss is used as the loss function. The word embeddings are set trainable and initialized with glove.840B.300d. The models with pre-trained contextualized word embeddings~\cite{rahman_integrating_2020} are hence excluded to ensure identical usage of external corpora. The optimizer is Adam for MulT and RMSprop for the other models, to be consistent with their source code implementations. A grid search for the best hyperparameters is performed for all models respectively, and the best performances on the test set are reported out of the total number of 50 searches. For each search, the model is trained for 100 epochs,and the model with the lowest validation loss is used to produce the test performance. 

The following hyparameters are grid-searched for our proposed quantum-inspired multimodal fusion (QMF) model: dimension of three modality inputs (after dimension reduction network) $t_{dim}, v_{dim},a_{dim} $ in $\{5,10,20\}$, local context lengths in powerset of $\{1,2,3,4\}$, number of eigenstates for the observable in $\{10,20,30,50,80\}$, size of the last hidden neuron in $\{16,32,48,64,80\}$, batch size in $\{32,64,96\}$ and learning rate in $\{0.001,0.002,0.005,0.008,0.01\}$. To commensurate with the main QMF framework, the models for pre-training the visual and acoustic phases involve $t_{dim}, v_{dim},a_{dim} \in \{5,10,20\}$, size of the last hidden neuron in $\{16,32,48,64,80\}$, batch size in $\{32,64,96\}$ and learning rate in $\{0.001,0.002,0.005,0.008,0.01\}$ as the grid parameters each.

A series of evaluation metrics are used in the experiment, in agreement with~\cite{bagher_zadeh_multimodal_2018}: 7-class accuracy, binary accuracy, F1 score, mean absolute error (MAE), and the correlation with human annotation. 

The models are implemented in python 3.6.8 and PyTorch 1.0.0. The experiments are run on a Linux server with Ubuntu 16.04.5 as the OS environment and 4 Nvidia Tesla V100 as the GPU devices. The source code of this work is available upon request.

%% file: results.tex
\section{Results and Discussion}
\subsection{Performance on Multimodal Sentiment Analysis}

The performance on CMU-MOSEI and CMU-MOSI is shown in Tab.~\ref{table:mosei} and~\ref{table:mosi} respectively. The bold values refer to the highest performance out of all the models for a specific metric. For each model, the percentage difference from the best score ($\%\Delta$) is shown in parentheses next to its absolute performance. The best hyperparameters for CMU-MOSEI are $t_{dim} = v_{dim} = a_{dim} = 10$, local context length $l=\{1,3\}$, the number of eigenstates $K=20$, last hidden layer size $h = 48$, batch size $bs = 32$, and learning rate $lr = 0.002$. The best settings for CMU-MOSI are $t_{dim} = v_{dim} = a_{dim} = 10$, $l=\{1,2\}$, $K=30$, $h = 24$, $bs = 32$, and $lr = 0.001$ respectively. 

Both tables indicate close results between our QMF and the best-performed models in the experiment. In particular, QMF obtains the best performance in MAE and Correlation and ranks second in binary accuracy and F1 value on CMU-MOSI. QMF is less competitive on CMU-MOSEI compared to other models, but it marginally underperforms the best model at a relative difference of less than 2.5$\%$ in all metrics. A significant performance discrepancy of over 2.5$\%$ between QMF and the best model is observed solely in 7-level accuracy on CMU-MOSI. We posit that it is because CMU-MOSI is a smaller dataset, and a minor increase in the number of wrong samples may lead to a non-negligible drop on the 7-level accuracy. In fact, the 7-level accuracy on CMU-MOSI has the greatest \textit{coefficient of variation} out of all metrics, suggesting low stability of this metric. On the larger dataset of CMU-MOSEI, QMF consistently outperforms MulT on all metrics, which was previously perceived as the best-performed model in this domain.

Another interesting finding is that different types of prior models, including advanced LSTM-based approaches, tensor-based models and seq2seq-based methods, are close to each other by effectiveness, and consistently outperform the simple EF-LSTM and LF-LSTM strategies. Even though similar trends have also been reported in the existing literature, the gaps observed in this experiment are much smaller. We conjecture that this is mainly because word embeddings are trained in this experiment while they were not set as trainable in previous models. Under a fixed word representation, the complexity of the neural structures may have an enormous impact on the representation capability of the model and hence influence performance. On the other hand, with trainable word embeddings, even a simple network structure may yield acceptable performance with a large number of training parameters in the embedding lookup table. This also explains why MulT was previously perceived as the best model but does not significantly outperform the remaining models in our experiment.

\begin{table}[t]
\begin{center}
\resizebox{\textwidth}{!}{
\begin{tabular}{c|c|c|c|c|c}%{|p{1.5cm}<{\centering}|p{1.5cm}<{\centering}|p{1.8cm}<{\centering}|r|}
 \hline \hline
%  $\Delta$
Model & Acc-7 & Acc-2 & F1 & MAE & Corr\\ \hline
\textbf{Vanilla LSTM} & & & & \\
EF-LSTM & 0.4753 (2.86\%) & 0.7921 (2.43\%) & 0.7895 (2.01\%) & 0.6560 (3.88\%) & 0.6268 (5.22\%)\\ 
LF-LSTM & 0.4719 (3.56\%) & 0.7911 (2.55\%) & 0.7855 (2.50\%) & 0.6669 (5.61\%) & 0.6102 (7.72\%) \\ \hline
\textbf{LSTM+} & & & & \\
MARN~\cite{zadeh_multi-attention_2018} & 0.4837 (1.14\%)  & 0.8090 (0.34\%) & 0.8014 (0.53\%) & \textbf{0.6310} & 0.6515 (1.48\%)\\ 
MFN~\cite{zadeh_memory_2018} &0.4448 (9.09\%) & 0.8031 (1.07\%) & 0.7925 (1.64\%) & 0.7044 (11.54\%) & 0.6562 (0.77\%)\\ \hline
% MCTN~\cite{sahay_multimodal_2018} &  &  & & &\\ \hline
\textbf{Tensor} & & & & \\
TFN~\cite{zadeh_tensor_2017} & \textbf{0.4893} & \textbf{0.8118} & \textbf{0.8079} & 0.6465 (2.38\%) & 0.6515 (1.48\%)\\
LMF~\cite{liu_efficient_2018} & 0.4824 (1.41\%)& 0.8064 (0.67\%) & 0.8057 (0.27\%) & 0.6358 (0.68\%)& \textbf{0.6613} \\ \hline
\textbf{Seq-to-Seq} & & & & \\
MulT~\cite{tsai_multimodal_2019} & 0.4590 (6.19\%) & 0.8022 (1.18\%) & 0.7951 (1.32\%) & 0.6980 (10.53\%) & 0.6511 (1.54\%) \\ \hline
\textbf{Ours} & & & & \\
QMF & 0.4788 (2.15\%)&  0.8069 (0.60\%)& 0.7977 (0.99\%) & 0.6399 (1.33\%) & 0.6575 (0.57\%)\\ \hline \hline
\end{tabular}
}
\caption{Effectiveness on CMU-MOSEI. The best scores out of all the models for a specific metric are in bold. The percentage difference from the best score ($\%\Delta$) is shown in parentheses next to the absolute performance of a model.}
\label{table:mosei}

\end{center}
\end{table}

\begin{table}[t]
\begin{center}

\resizebox{\textwidth}{!}{
\begin{tabular}{c|c|c|c|c|c}%{|p{1.5cm}<{\centering}|p{1.5cm}<{\centering}|p{1.8cm}<{\centering}|r|}
 \hline \hline
%  $\Delta$
Model & Acc-7 & Acc-2 & F1 & MAE & Corr\\ \hline
\textbf{Vanilla LSTM} & & & & \\
EF-LSTM & 0.3323 (9.90\%) & 0.7770 (2.90\%) & 0.7772 (2.60\%) & 0.9675 (5.78\%)& 0.6504 (6.54\%)\\ 
LF-LSTM & 0.3178 (13.83\%) & 0.7711 (3.63\%) & 0.7702 (3.48\%) & 0.9768 (6.80\%) & 0.6381 (8.31\%)\\ \hline
\textbf{LSTM+} & & & & \\
MARN~\cite{zadeh_multi-attention_2018} & 0.3294 (10.68\%) & 0.7959 (0.54\%) & 0.7955 (0.31\%) & 0.9576  (4.70\%) & 0.6739 (3.16\%) \\ 
MFN~\cite{zadeh_memory_2018} & 0.3236 (12.26\%) & 0.7851 (1.89\%) & 0.7838 (1.78\%)& 0.9684 (5.88\%) &0.6380 (8.32\%)\\ \hline
\textbf{Tensor} & & & & \\
% MCTN~\cite{sahay_multimodal_2018} &  &  & & &\\ \hline
TFN~\cite{zadeh_tensor_2017} & 0.3586 (2.77\%) & 0.7784 (2.72\%) & 0.7785 (2.44\%) & 0.9642 (5.42\%) & 0.6591 (5.29\%)\\
LMF~\cite{liu_efficient_2018} & \textbf{0.3688}  & 0.7872 (1.62\%) & 0.7871 (1.37\%)& 0.9409 (2.88\%) & 0.6595 (5.23\%)\\ \hline
\textbf{Seq-to-Seq} & & & & \\
MulT~\cite{tsai_multimodal_2019} & 0.3528 (4.34\%)  & \textbf{0.8002} & \textbf{0.7980} & 0.9407 (2.86\%) & 0.6911 (0.69\%)\\ \hline
\textbf{Ours} & & & & \\
QMF & 0.3353 (9.08\%) & 0.7974 (0.35\%) & 0.7962 (0.23\%) & \textbf{0.9146} & \textbf{0.6959}\\  \hline \hline

\hline
\end{tabular}
}
\caption{Effectiveness on CMU-MOSEI. The best scores out of all the models for a specific metric are in bold. The percentage difference from the best score ($\%\Delta$) is shown in parentheses next to the absolute performance of a model.}
\label{table:mosi}

\end{center}
\end{table}

\subsection{Ablation Study}

In order to examine the influence of each component in the proposed model, an ablation study is conducted on the larger of the two datasets, CMU-MOSEI. Based on the best settings, changes are made only on the respective component, so that the performance difference is a reliable indicator of the impact of the element.

To validate the effectiveness of the modulus-argument assignment of complex-valued embedding, we replace the complex-valued components with their real counterparts. However, simple removal of the arguments will lead to a decrease in parameter scale and may bias the results. In order to eliminate this effect, the real-valued network QMF-real contains doubled dimensions $t_{dim} = 20,v_{dim} = 20, a_{dim} =20$ for unimodal inputs and twice the number of sentiment-related aspects $K = 40$. 

A special strategy is introduced to initialize the arguments of three modalities. To check whether it positively affects the model performance, we re-train the same model with randomly initialized arguments (i.e., QMF-rand-init) and compare its performance with the original QMF model.

Another crucial network unit is the local-mixture strategy, where the density matrices of local contexts are extracted and fed to the measurement. To justify the use of this component, we run a model with a global mixture of all words in the sentence (i.e., QMF-global-mixture), with the other setting unchanged.

Finally, after the measurement results are outputted, a row-wise max-pooling is conducted to identify the most representative context for each sentiment-related aspect (i.e., eigenstate). We contrast this strategy with the row-wise average-pooling (QMF-average-pool), which uses the average probability of all local contexts to represent the sentence feature with respect to a particular aspect.

\begin{table}[t]
\begin{center}

\resizebox{0.8\textwidth}{!}{
\begin{tabular}{l|l|l|l|l|l}%{|p{1.5cm}<{\centering}|p{1.5cm}<{\centering}|p{1.8cm}<{\centering}|r|}
 \hline \hline
%  $\Delta$
Models & Acc-7 & Acc-2 & F1 & MAE & Corr\\ \hline
QMF & 0.4788 & 0.8069 & 0.7977 & 0.6399 & 0.6575\\ \hline
QMF-real & 0.4241 & 0.7301& 0.7320 & 0.7641 & 0.4682\\ \hline
QMF-rand-init & 0.4221 & 0.7172 & 0.7278 & 0.7583 & 0.5332\\ \hline
QMF-global-mixture & 0.4324  & 0.7237 & 0.7244 & 0.7671  & 0.4215 \\ \hline
QMF-average-pool & 0.4208 & 0.7325 & 0.7401 & 0.7102 & 0.5542 \\ \hline
\hline
\end{tabular}
}
\caption{Ablation Study on CMU-MOSEI.}
\label{table:ablation}

\end{center}
\end{table}

As shown in Tab.~\ref{table:ablation}, a notable drop in performance is observed for all QMF variants. This illustrates the usefulness of complex-valued components, arguments initialization strategies, the local mixture strategy, as well as the max-pooling for measurement results. In particular, the discrepancy with QMF-real empirically suggests that the complex values in the components are not merely a doubling of parameters, but bring about a meaningful combination of the respective features for the modulus and argument parts (which agrees with Sec.~\ref{sec:embedding}) that leads to a positive performance gain for the whole model.

\subsection{Interpretation of Multimodal Decision}\label{sec:inter_analysis}
Our proposed model captures inter-modal interactions on the decision level, viewing the multimodal sentiment judgment as an entanglement of unimodal decisions. In order to understand the \textit{entangled sentiment decision}, we disentangle the best QMF model on CMU-MOSEI by looking into the decisions on unimodal and bimodal data implicitly encoded in the trimodal sentiment analysis. The particular approach to make this available is introduced in Section~\ref{sec:interpretation}. 

\begin{table}
\begin{center}
\resizebox{0.8\textwidth}{!}{
\begin{tabular}{l|l|l|l|l|l}%{|p{1.5cm}<{\centering}|p{1.5cm}<{\centering}|p{1.8cm}<{\centering}|r|}
 \hline \hline
%  $\Delta$
Models & Acc-7 & Acc-2 & F1 & MAE & Corr \\ \hline
QMF-trimodal & 0.4788 & 0.8069 & 0.7977 & 0.6399 & 0.6575\\ \hline
QMF-textual & 0.3644 & 0.7629 & 0.7064 & 1.1789& 0.4949\\ \hline
QMF-visual &0.2893 & 0.4135 & 0.1299 & 0.9400 & 0.1608 \\ \hline
QMF-acoustic &0.2897 & 0.4137 & 0.1301 & 0.9623 & 0.0313\\ \hline
QMF-textual+visual & 0.3923 & 0.7973 & 0.7780 & 0.7800 & 0.5796 \\ \hline
QMF-textual+acoustic &0.3955 & 0.7971 & 0.7748 & 0.7305 & 0.5509\\ \hline
QMF-visual+acoustic & 0.2053 & 0.2897 & 0.1301 & 1.0731 & 0.2073\\ \hline

\hline
\end{tabular}
}
\caption{Unimodal and bimodal sentiment classification result on CMU-MOSEI, entailed by the best-performed QMF learned by the whole CMU-MOSEI data.}
\label{table:interpretation}

\end{center}
\end{table}

Tab.~\ref{table:interpretation} shows the sentiment prediction results based on all unimodal and combinations of bimodal features of MOSEI. The results show that the QMF best predicts the sentiment based on three modalities. When QMF is used to predict sentiment based on unimodal data, it is able to give a reasonably accurate prediction for textual features, but barely able to provide any predictions for visual and acoustic modalities. However, the QMF is able to give better judgments when combining visual or acoustic features with textual features, as can be seen from the gradually increased performances in both textual $\rightarrow$ textual+visual $\rightarrow$ textual+visual+acoustic and textual $\rightarrow$ textual+acoustic $\rightarrow$ textual+visual+acoustic paths.

The results above indicate that the textual modality plays a predominant role in determining the sentiment, while visual and acoustic modalities are less relevant to the sentence sentiment. This finding is consistent with the prior work in this field. Furthermore, even if the textual modality carries the majority of sentiment-related information, complementary information is extracted from the visual and acoustic modalities to boost the sentiment prediction capability of textual modality. This process leads to reduced abilities for the visual and acoustics to predict sentiment independently.

\paragraph{Summary}
Overall, our proposed QMF is comparable to SOTA baseline models on both CMU-MOSEI and CMU-MOSI. The effectiveness is further supported in an ablation study, where the introduction of complex-valued representations, mixture of local contexts, global-mixture of measurement results, and the initialization method all positively influence the model effectiveness. Alternatively, we present the model predictions on unimodal and bimodal features of CMU-MOSEI to achieve an understanding of how unimodal decisions compose an entangled multimodal decision.

%% file: conclusion.tex
\section{Conclusion and Future Work}

We have built a novel quantum-inspired framework for multimodal sentiment analysis. The framework borrows quantum concepts to explicitly model intra-modal interactions on the feature level and inter-modal interactions on the decision level. A neural network with complex-valued components is built to learn both interactions in an end-to-end supervised way. In addition to obtaining comparable performance to state-of-the-art models, we also contribute an interpretation approach to facilitate understanding of multimodal interactions from both quantum and classical perspectives.

We also note that the quality of the extracted visual and acoustic features is not high. It would be a promising direction to extract clean and sentiment-sensitive features from non-textual modalities so that the value of models in this field could be appropriately judged. On the other hand, the inconsistency with quantum theory shall be handled for the measurement component. If a legal observable with an orthogonal basis as eigenstates and concrete meanings for observed values could be incorporated into the end-to-end network, the model would be more interpretable and effective.

%% file: acknowledgements.tex
\section*{Acknowledgements}

This study is supported by the Quantum Information Access and Retrieval Theory (QUARTZ) project, which has received funding from the European Union's Horizon 2020 research and innovation programme under the Marie Sk\l{}odowska-Curie grant agreement No. 721321. We would like to thank Prof. Jakob Simenson for his valuable feedback.

%% file: main.bbl
\begin{thebibliography}{10}
\expandafter\ifx\csname url\endcsname\relax
  \def\url#1{\texttt{#1}}\fi
\expandafter\ifx\csname urlprefix\endcsname\relax\def\urlprefix{URL }\fi
\expandafter\ifx\csname href\endcsname\relax
  \def\href#1#2{#2} \def\path#1{#1}\fi

\bibitem{zadeh_tensor_2017}
A.~Zadeh, M.~Chen, S.~Poria, E.~Cambria, L.-P. Morency,
  \href{https://www.aclweb.org/anthology/D17-1115/}{Tensor {Fusion} {Network}
  for {Multimodal} {Sentiment} {Analysis}}, in: Proceedings of the 55th
  {Annual} {Meeting} of the {Association} for {Computational} {Linguistics}
  ({Volume} 1: {Long} {Papers}), Association for Computational Linguistics,
  2017, pp. 1103--1114.
\newblock \href {http://dx.doi.org/10.18653/v1/D17-1115}
  {\path{doi:10.18653/v1/D17-1115}}.
\newline\urlprefix\url{https://www.aclweb.org/anthology/D17-1115/}

\bibitem{liu_efficient_2018}
Z.~Liu, Y.~Shen, V.~B. Lakshminarasimhan, P.~P. Liang, A.~Bagher~Zadeh, L.-P.
  Morency, \href{http://www.aclweb.org/anthology/P18-1209}{Efficient {Low}-rank
  {Multimodal} {Fusion} {With} {Modality}-{Specific} {Factors}}, in:
  Proceedings of the 56th {Annual} {Meeting} of the {Association} for
  {Computational} {Linguistics} ({Volume} 1: {Long} {Papers}), Association for
  Computational Linguistics, Melbourne, Australia, 2018, pp. 2247--2256.
\newline\urlprefix\url{http://www.aclweb.org/anthology/P18-1209}

\bibitem{liang_multimodal_2018}
P.~P. Liang, Z.~Liu, A.~Bagher~Zadeh, L.-P. Morency,
  \href{https://www.aclweb.org/anthology/D18-1014}{Multimodal {Language}
  {Analysis} with {Recurrent} {Multistage} {Fusion}}, in: Proceedings of the
  2018 {Conference} on {Empirical} {Methods} in {Natural} {Language}
  {Processing}, Association for Computational Linguistics, Brussels, Belgium,
  2018, pp. 150--161.
\newblock \href {http://dx.doi.org/10.18653/v1/D18-1014}
  {\path{doi:10.18653/v1/D18-1014}}.
\newline\urlprefix\url{https://www.aclweb.org/anthology/D18-1014}

\bibitem{zadeh_multi-attention_2018}
A.~Zadeh, P.~P. Liang, S.~Poria, P.~Vij, E.~Cambria, L.-P. Morency,
  \href{https://aaai.org/ocs/index.php/AAAI/AAAI18/paper/view/17390}{Multi-attention
  {Recurrent} {Network} for {Human} {Communication} {Comprehension}}, in:
  Thirty-{Second} {AAAI} {Conference} on {Artificial} {Intelligence}, 2018, pp.
  5643--5649.
\newline\urlprefix\url{https://aaai.org/ocs/index.php/AAAI/AAAI18/paper/view/17390}

\bibitem{zadeh_memory_2018}
A.~Zadeh, P.~P. Liang, N.~Mazumder, S.~Poria, E.~Cambria, L.-P. Morency,
  \href{https://aaai.org/ocs/index.php/AAAI/AAAI18/paper/view/17341}{Memory
  {Fusion} {Network} for {Multi}-view {Sequential} {Learning}}, in:
  Thirty-{Second} {AAAI} {Conference} on {Artificial} {Intelligence}, 2018, pp.
  5634--5641.
\newline\urlprefix\url{https://aaai.org/ocs/index.php/AAAI/AAAI18/paper/view/17341}

\bibitem{bagher_zadeh_multimodal_2018}
A.~Bagher~Zadeh, P.~P. Liang, S.~Poria, E.~Cambria, L.-P. Morency,
  \href{http://www.aclweb.org/anthology/P18-1208}{Multimodal {Language}
  {Analysis} in the {Wild}: {CMU}-{MOSEI} {Dataset} and {Interpretable}
  {Dynamic} {Fusion} {Graph}}, in: Proceedings of the 56th {Annual} {Meeting}
  of the {Association} for {Computational} {Linguistics} ({Volume} 1: {Long}
  {Papers}), Association for Computational Linguistics, Melbourne, Australia,
  2018, pp. 2236--2246.
\newline\urlprefix\url{http://www.aclweb.org/anthology/P18-1208}

\bibitem{pham_found_2019}
H.~Pham, P.~P. Liang, T.~Manzini, L.-P. Morency, B.~Póczos,
  \href{https://aaai.org/ojs/index.php/AAAI/article/view/4666}{Found in
  {Translation}: {Learning} {Robust} {Joint} {Representations} by {Cyclic}
  {Translations} between {Modalities}}, 1 33 (2019) 6892--6899.
\newblock \href {http://dx.doi.org/10.1609/aaai.v33i01.33016892}
  {\path{doi:10.1609/aaai.v33i01.33016892}}.
\newline\urlprefix\url{https://aaai.org/ojs/index.php/AAAI/article/view/4666}

\bibitem{tsai_multimodal_2019}
Y.-H.~H. Tsai, S.~Bai, P.~P. Liang, J.~Z. Kolter, L.-P. Morency,
  R.~Salakhutdinov, \href{https://www.aclweb.org/anthology/P19-1656}{Multimodal
  {Transformer} for {Unaligned} {Multimodal} {Language} {Sequences}}, in:
  Proceedings of the 57th {Annual} {Meeting} of the {Association} for
  {Computational} {Linguistics}, Association for Computational Linguistics,
  Florence, Italy, 2019, pp. 6558--6569.
\newline\urlprefix\url{https://www.aclweb.org/anthology/P19-1656}

\bibitem{barezi_modality-based_2019}
E.~J. Barezi, P.~Fung,
  \href{https://www.aclweb.org/anthology/W19-4331/}{Modality-based
  {Factorization} for {Multimodal} {Fusion}}, in: Proceedings of the 4th
  Workshop on Representation Learning for NLP (RepL4NLP-2019), 2019, pp.
  260--269.
\newline\urlprefix\url{https://www.aclweb.org/anthology/W19-4331/}

\bibitem{liang_learning_2019}
P.~P. Liang, Z.~Liu, Y.-H.~H. Tsai, Q.~Zhao, R.~Salakhutdinov, L.-P. Morency,
  \href{https://www.aclweb.org/anthology/P19-1152}{Learning {Representations}
  from {Imperfect} {Time} {Series} {Data} via {Tensor} {Rank}
  {Regularization}}, in: Proceedings of the 57th {Annual} {Meeting} of the
  {Association} for {Computational} {Linguistics}, Association for
  Computational Linguistics, Florence, Italy, 2019, pp. 1569--1576.
\newline\urlprefix\url{https://www.aclweb.org/anthology/P19-1152}

\bibitem{mai_divide_2019}
S.~Mai, H.~Hu, S.~Xing,
  \href{https://www.aclweb.org/anthology/P19-1046}{Divide, {Conquer} and
  {Combine}: {Hierarchical} {Feature} {Fusion} {Network} with {Local} and
  {Global} {Perspectives} for {Multimodal} {Affective} {Computing}}, in:
  Proceedings of the 57th {Annual} {Meeting} of the {Association} for
  {Computational} {Linguistics}, Association for Computational Linguistics,
  Florence, Italy, 2019, pp. 481--492.
\newline\urlprefix\url{https://www.aclweb.org/anthology/P19-1046}

\bibitem{wang_words_2019}
Y.~Wang, Y.~Shen, Z.~Liu, P.~P. Liang, A.~Zadeh, L.-P. Morency,
  \href{https://aaai.org/ojs/index.php/AAAI/article/view/4706}{Words {Can}
  {Shift}: {Dynamically} {Adjusting} {Word} {Representations} {Using}
  {Nonverbal} {Behaviors}}, 1 33 (2019) 7216--7223.
\newblock \href {http://dx.doi.org/10.1609/aaai.v33i01.33017216}
  {\path{doi:10.1609/aaai.v33i01.33017216}}.
\newline\urlprefix\url{https://aaai.org/ojs/index.php/AAAI/article/view/4706}

\bibitem{tsai_learning_2019}
Y.-H.~H. Tsai, P.~P. Liang, A.~Zadeh, L.-P. Morency, R.~Salakhutdinov,
  {LEARNING} {FACTORIZED} {MULTIMODAL} {REPRESENTATIONS} (2019) 20.

\bibitem{rahman_integrating_2020}
W.~Rahman, M.~K. Hasan, S.~Lee, A.~Bagher~Zadeh, C.~Mao, L.-P. Morency,
  E.~Hoque,
  \href{https://www.aclweb.org/anthology/2020.acl-main.214}{Integrating
  {Multimodal} {Information} in {Large} {Pretrained} {Transformers}}, in:
  Proceedings of the 58th {Annual} {Meeting} of the {Association} for
  {Computational} {Linguistics}, Association for Computational Linguistics,
  Online, 2020, pp. 2359--2369.
\newline\urlprefix\url{https://www.aclweb.org/anthology/2020.acl-main.214}

\bibitem{baltrusaitis_multimodal_2017}
T.~Baltrušaitis, C.~Ahuja, L.-P. Morency,
  \href{http://arxiv.org/abs/1705.09406}{Multimodal {Machine} {Learning}: {A}
  {Survey} and {Taxonomy}}, arXiv:1705.09406 [cs]ArXiv: 1705.09406.
\newline\urlprefix\url{http://arxiv.org/abs/1705.09406}

\bibitem{lipton_mythos_2018}
Z.~C. Lipton, \href{http://doi.acm.org/10.1145/3236386.3241340}{The {Mythos} of
  {Model} {Interpretability}}, Queue 16~(3) (2018) 30:31--30:57.
\newblock \href {http://dx.doi.org/10.1145/3236386.3241340}
  {\path{doi:10.1145/3236386.3241340}}.
\newline\urlprefix\url{http://doi.acm.org/10.1145/3236386.3241340}

\bibitem{holzinger_current_2018}
A.~Holzinger, P.~Kieseberg, E.~Weippl, A.~M. Tjoa, Current {Advances}, {Trends}
  and {Challenges} of {Machine} {Learning} and {Knowledge} {Extraction}: {From}
  {Machine} {Learning} to {Explainable} {AI}, in: A.~Holzinger, P.~Kieseberg,
  A.~M. Tjoa, E.~Weippl (Eds.), Machine {Learning} and {Knowledge}
  {Extraction}, Springer International Publishing, Cham, 2018, pp. 1--8.

\bibitem{busemeyer_quantum_2012}
J.~R. Busemeyer, P.~D. Bruza, Quantum {Models} of {Cognition} and {Decision},
  1st Edition, Cambridge University Press, New York, NY, USA, 2012.

\bibitem{sordoni_modeling_2013}
A.~Sordoni, J.~He, J.-Y. Nie,
  \href{http://doi.acm.org/10.1145/2505515.2507854}{Modeling {Latent} {Topic}
  {Interactions} {Using} {Quantum} {Interference} for {Information}
  {Retrieval}}, in: Proceedings of the 22Nd {ACM} {International} {Conference}
  on {Information} \& {Knowledge} {Management}, {CIKM} '13, ACM, New York, NY,
  USA, 2013, pp. 1197--1200.
\newblock \href {http://dx.doi.org/10.1145/2505515.2507854}
  {\path{doi:10.1145/2505515.2507854}}.
\newline\urlprefix\url{http://doi.acm.org/10.1145/2505515.2507854}

\bibitem{li_quantum-inspired_2018}
Q.~Li, S.~Uprety, B.~Wang, D.~Song,
  \href{https://aclanthology.info/papers/W18-3006/w18-3006}{Quantum-{Inspired}
  {Complex} {Word} {Embedding}}, Proceedings of The Third Workshop on
  Representation Learning for NLP (2018) 50--57.
\newline\urlprefix\url{https://aclanthology.info/papers/W18-3006/w18-3006}

\bibitem{wang_semantic_2019}
B.~Wang, Q.~Li, M.~Melucci, D.~Song,
  \href{http://doi.acm.org/10.1145/3308558.3313516}{Semantic {Hilbert} {Space}
  for {Text} {Representation} {Learning}}, in: The {World} {Wide} {Web}
  {Conference}, {WWW} '19, ACM, New York, NY, USA, 2019, pp. 3293--3299,
  event-place: San Francisco, CA, USA.
\newblock \href {http://dx.doi.org/10.1145/3308558.3313516}
  {\path{doi:10.1145/3308558.3313516}}.
\newline\urlprefix\url{http://doi.acm.org/10.1145/3308558.3313516}

\bibitem{li_cnm:_2019}
Q.~Li, B.~Wang, M.~Melucci,
  \href{https://www.aclweb.org/anthology/N19-1420}{{CNM}: {An} {Interpretable}
  {Complex}-valued {Network} for {Matching}}, in: Proceedings of the 2019
  {Conference} of the {North} {American} {Chapter} of the {Association} for
  {Computational} {Linguistics}: {Human} {Language} {Technologies}, {Volume} 1
  ({Long} and {Short} {Papers}), Association for Computational Linguistics,
  Minneapolis, Minnesota, 2019, pp. 4139--4148.
\newline\urlprefix\url{https://www.aclweb.org/anthology/N19-1420}

\bibitem{zhang_quantum-inspired_2018}
Y.~Zhang, D.~Song, P.~Zhang, P.~Wang, J.~Li, X.~Li, B.~Wang,
  \href{http://www.sciencedirect.com/science/article/pii/S0304397518302639}{A
  quantum-inspired multimodal sentiment analysis framework}, Theoretical
  Computer Science\href {http://dx.doi.org/10.1016/j.tcs.2018.04.029}
  {\path{doi:10.1016/j.tcs.2018.04.029}}.
\newline\urlprefix\url{http://www.sciencedirect.com/science/article/pii/S0304397518302639}

\bibitem{gkoumas_investigating_2018}
D.~Gkoumas, S.~Uprety, D.~Song, Investigating non-classical correlations
  between decision fused multi-modal documents, Quantum Interaction 2018.

\bibitem{zadeh_mosi:_2016}
A.~Zadeh, R.~Zellers, E.~Pincus, L.-P. Morency,
  \href{http://arxiv.org/abs/1606.06259}{{MOSI}: {Multimodal} {Corpus} of
  {Sentiment} {Intensity} and {Subjectivity} {Analysis} in {Online} {Opinion}
  {Videos}}, arXiv:1606.06259 [cs]ArXiv: 1606.06259.
\newline\urlprefix\url{http://arxiv.org/abs/1606.06259}

\bibitem{gleason1957measures}
A.~M. Gleason, Measures on the closed subspaces of a hilbert space, Journal of
  mathematics and mechanics (1957) 885--893.

\bibitem{melucci_introduction_2015}
M.~Melucci,
  \href{http://link.springer.com/10.1007/978-3-662-48313-8}{Introduction to
  {Information} {Retrieval} and {Quantum} {Mechanics}}, Vol.~35 of The
  {Information} {Retrieval} {Series}, Springer Berlin Heidelberg, Berlin,
  Heidelberg, 2015.
\newblock \href {http://dx.doi.org/10.1007/978-3-662-48313-8}
  {\path{doi:10.1007/978-3-662-48313-8}}.
\newline\urlprefix\url{http://link.springer.com/10.1007/978-3-662-48313-8}

\bibitem{Zeilinger10}
A.~Zeilinger, Dance of the Photons: From Einstein to Quantum Teleportation,
  Farrar, Straus and Giroux, 2010.

\bibitem{nielsen_quantum_2011}
M.~A. Nielsen, I.~L. Chuang, Quantum {Computation} and {Quantum} {Information}:
  10th {Anniversary} {Edition}, 10th Edition, Cambridge University Press, New
  York, NY, USA, 2011.

\bibitem{shutova_black_2016}
E.~Shutova, D.~Kiela, J.~Maillard,
  \href{http://aclweb.org/anthology/N16-1020}{Black {Holes} and {White}
  {Rabbits}: {Metaphor} {Identification} with {Visual} {Features}}, in:
  Proceedings of the 2016 {Conference} of the {North} {American} {Chapter} of
  the {Association} for {Computational} {Linguistics}: {Human} {Language}
  {Technologies}, Association for Computational Linguistics, 2016, pp.
  160--170, event-place: San Diego, California.
\newblock \href {http://dx.doi.org/10.18653/v1/N16-1020}
  {\path{doi:10.18653/v1/N16-1020}}.
\newline\urlprefix\url{http://aclweb.org/anthology/N16-1020}

\bibitem{morvant_majority_2014}
E.~Morvant, A.~Habrard, S.~Ayache,
  \href{http://arxiv.org/abs/1404.7796}{Majority {Vote} of {Diverse}
  {Classifiers} for {Late} {Fusion}}, arXiv:1404.7796 [cs, stat]ArXiv:
  1404.7796.
\newline\urlprefix\url{http://arxiv.org/abs/1404.7796}

\bibitem{evangelopoulos_multimodal_2013}
G.~{Evangelopoulos}, A.~{Zlatintsi}, A.~{Potamianos}, P.~{Maragos},
  K.~{Rapantzikos}, G.~{Skoumas}, Y.~{Avrithis}, Multimodal saliency and fusion
  for movie summarization based on aural, visual, and textual attention, IEEE
  Transactions on Multimedia 15~(7) (2013) 1553--1568.

\bibitem{glodek_multiple_2011}
M.~Glodek, S.~Tschechne, G.~Layher, M.~Schels, T.~Brosch, S.~Scherer,
  M.~Kächele, M.~Schmidt, H.~Neumann, G.~Palm, F.~Schwenker, Multiple
  {Classifier} {Systems} for the {Classification} of {Audio}-{Visual}
  {Emotional} {States}, in: S.~D'Mello, A.~Graesser, B.~Schuller, J.-C.
  Martin (Eds.), Affective {Computing} and {Intelligent} {Interaction}, Lecture
  {Notes} in {Computer} {Science}, Springer Berlin Heidelberg, 2011, pp.
  359--368.

\bibitem{vaswani_attention_2017}
A.~Vaswani, N.~Shazeer, N.~Parmar, J.~Uszkoreit, L.~Jones, A.~N. Gomez,
  Å.~Kaiser, I.~Polosukhin,
  \href{http://papers.nips.cc/paper/7181-attention-is-all-you-need.pdf}{Attention
  is {All} you {Need}}, in: I.~Guyon, U.~V. Luxburg, S.~Bengio, H.~Wallach,
  R.~Fergus, S.~Vishwanathan, R.~Garnett (Eds.), Advances in {Neural}
  {Information} {Processing} {Systems} 30, Curran Associates, Inc., 2017, pp.
  5998--6008.
\newline\urlprefix\url{http://papers.nips.cc/paper/7181-attention-is-all-you-need.pdf}

\bibitem{devlin_bert:_2019}
J.~Devlin, M.-W. Chang, K.~Lee, K.~Toutanova,
  \href{https://www.aclweb.org/anthology/N19-1423}{{BERT}: {Pre}-training of
  {Deep} {Bidirectional} {Transformers} for {Language} {Understanding}}, in:
  Proceedings of the 2019 {Conference} of the {North} {American} {Chapter} of
  the {Association} for {Computational} {Linguistics}: {Human} {Language}
  {Technologies}, {Volume} 1 ({Long} and {Short} {Papers}), Association for
  Computational Linguistics, Minneapolis, Minnesota, 2019, pp. 4171--4186.
\newblock \href {http://dx.doi.org/10.18653/v1/N19-1423}
  {\path{doi:10.18653/v1/N19-1423}}.
\newline\urlprefix\url{https://www.aclweb.org/anthology/N19-1423}

\bibitem{yang_xlnet_2019}
Z.~Yang, Z.~Dai, Y.~Yang, J.~Carbonell, R.~R. Salakhutdinov, Q.~V. Le,
  \href{http://papers.nips.cc/paper/8812-xlnet-generalized-autoregressive-pretraining-for-language-understanding.pdf}{{XLNet}:
  {Generalized} {Autoregressive} {Pretraining} for {Language} {Understanding}},
  in: H.~Wallach, H.~Larochelle, A.~Beygelzimer, F.~d. Alché-Buc, E.~Fox,
  R.~Garnett (Eds.), Advances in {Neural} {Information} {Processing} {Systems}
  32, Curran Associates, Inc., 2019, pp. 5753--5763.
\newline\urlprefix\url{http://papers.nips.cc/paper/8812-xlnet-generalized-autoregressive-pretraining-for-language-understanding.pdf}

\bibitem{chaturvedi_fuzzy_2019}
I.~Chaturvedi, R.~Satapathy, S.~Cavallari, E.~Cambria,
  \href{http://www.sciencedirect.com/science/article/pii/S0167865519301394}{Fuzzy
  commonsense reasoning for multimodal sentiment analysis}, Pattern Recognition
  Letters 125 (2019) 264--270.
\newblock \href {http://dx.doi.org/10.1016/j.patrec.2019.04.024}
  {\path{doi:10.1016/j.patrec.2019.04.024}}.
\newline\urlprefix\url{http://www.sciencedirect.com/science/article/pii/S0167865519301394}

\bibitem{cambria_sentic_2013}
E.~Cambria, N.~Howard, J.~Hsu, A.~Hussain,
  \href{http://ieeexplore.ieee.org/document/6613272/}{Sentic blending:
  {Scalable} multimodal fusion for the continuous interpretation of semantics
  and sentics}, in: 2013 {IEEE} {Symposium} on {Computational} {Intelligence}
  for {Human}-like {Intelligence} ({CIHLI}), IEEE, Singapore, Singapore, 2013,
  pp. 108--117.
\newblock \href {http://dx.doi.org/10.1109/CIHLI.2013.6613272}
  {\path{doi:10.1109/CIHLI.2013.6613272}}.
\newline\urlprefix\url{http://ieeexplore.ieee.org/document/6613272/}

\bibitem{wang_tensor_2010}
J.~Wang, D.~Song, K.~Leszek,
  \href{https://www.aaai.org/ocs/index.php/FSS/FSS10/paper/view/2297}{Tensor
  {Product} of {Correlated} {Textual} and {Visual} {Features}: {A} {Quantum}
  {Theory} {Inspired} {Image} {Retrieval} {Framework}}, AAAI Fall Symposium
  Series.
\newline\urlprefix\url{https://www.aaai.org/ocs/index.php/FSS/FSS10/paper/view/2297}

\bibitem{rijsbergen_geometry_2004}
C.~J.~v. Rijsbergen, The {Geometry} of {Information} {Retrieval}, Cambridge
  University Press, New York, NY, USA, 2004.

\bibitem{zuccon_using_2010}
G.~Zuccon, L.~Azzopardi,
  \href{http://dx.doi.org/10.1007/978-3-642-12275-0-32}{Using the {Quantum}
  {Probability} {Ranking} {Principle} to {Rank} {Interdependent} {Documents}},
  in: Proceedings of the 32Nd {European} {Conference} on {Advances} in
  {Information} {Retrieval}, {ECIR}'2010, Springer-Verlag, Berlin, Heidelberg,
  2010, pp. 357--369.
\newblock \href {http://dx.doi.org/10.1007/978-3-642-12275-0-32}
  {\path{doi:10.1007/978-3-642-12275-0-32}}.
\newline\urlprefix\url{http://dx.doi.org/10.1007/978-3-642-12275-0-32}

\bibitem{zhang_quantum_2016}
P.~Zhang, J.~Li, B.~Wang, X.~Zhao, D.~Song, Y.~Hou, M.~Melucci,
  \href{http://www.mdpi.com/1099-4300/18/4/146}{A {Quantum} {Query} {Expansion}
  {Approach} for {Session} {Search}}, Entropy 18~(4) (2016) 146.
\newblock \href {http://dx.doi.org/10.3390/e18040146}
  {\path{doi:10.3390/e18040146}}.
\newline\urlprefix\url{http://www.mdpi.com/1099-4300/18/4/146}

\bibitem{melucci_towards_2008}
M.~Melucci, Towards modeling implicit feedback with quantum entanglement,
  Quantum Interaction 2008.

\bibitem{wang_exploration_2016}
B.~Wang, P.~Zhang, J.~Li, D.~Song, Y.~Hou, Z.~Shang,
  \href{http://www.mdpi.com/1099-4300/18/4/144}{Exploration of {Quantum}
  {Interference} in {Document} {Relevance} {Judgement} {Discrepancy}}, Entropy
  18~(4) (2016) 144.
\newblock \href {http://dx.doi.org/10.3390/e18040144}
  {\path{doi:10.3390/e18040144}}.
\newline\urlprefix\url{http://www.mdpi.com/1099-4300/18/4/144}

\bibitem{sordoni_modeling_2013_2}
A.~Sordoni, J.-Y. Nie, Y.~Bengio,
  \href{http://doi.acm.org/10.1145/2484028.2484098}{Modeling {Term}
  {Dependencies} with {Quantum} {Language} {Models} for {IR}}, in: Proceedings
  of the 36th {International} {ACM} {SIGIR} {Conference} on {Research} and
  {Development} in {Information} {Retrieval}, {SIGIR} '13, ACM, New York, NY,
  USA, 2013, pp. 653--662.
\newblock \href {http://dx.doi.org/10.1145/2484028.2484098}
  {\path{doi:10.1145/2484028.2484098}}.
\newline\urlprefix\url{http://doi.acm.org/10.1145/2484028.2484098}

\bibitem{xie_modeling_2015}
M.~Xie, Y.~Hou, P.~Zhang, J.~Li, W.~Li, D.~Song,
  \href{http://dl.acm.org/citation.cfm?id=2832415.2832439}{Modeling {Quantum}
  {Entanglements} in {Quantum} {Language} {Models}}, in: Proceedings of the
  24th {International} {Conference} on {Artificial} {Intelligence}, {IJCAI}'15,
  AAAI Press, Buenos Aires, Argentina, 2015, pp. 1362--1368.
\newline\urlprefix\url{http://dl.acm.org/citation.cfm?id=2832415.2832439}

\bibitem{li_modeling_2015}
Q.~Li, J.~Li, P.~Zhang, D.~Song,
  \href{http://doi.acm.org/10.1145/2766462.2767819}{Modeling {Multi}-query
  {Retrieval} {Tasks} {Using} {Density} {Matrix} {Transformation}}, in:
  Proceedings of the 38th {International} {ACM} {SIGIR} {Conference} on
  {Research} and {Development} in {Information} {Retrieval}, {SIGIR} '15, ACM,
  New York, NY, USA, 2015, pp. 871--874.
\newblock \href {http://dx.doi.org/10.1145/2766462.2767819}
  {\path{doi:10.1145/2766462.2767819}}.
\newline\urlprefix\url{http://doi.acm.org/10.1145/2766462.2767819}

\bibitem{balkir_distributional_2016}
E.~Balkir, M.~Sadrzadeh, B.~Coecke, Distributional {Sentence} {Entailment}
  {Using} {Density} {Matrices}, in: M.~T. Hajiaghayi, M.~R. Mousavi (Eds.),
  Topics in {Theoretical} {Computer} {Science}, Lecture {Notes} in {Computer}
  {Science}, Springer International Publishing, 2016, pp. 1--22.

\bibitem{blacoe_quantum-theoretic_2013}
W.~Blacoe, E.~Kashefi, M.~Lapata,
  \href{http://www.aclweb.org/anthology/N13-1105}{A {Quantum}-{Theoretic}
  {Approach} to {Distributional} {Semantics}}, in: Proceedings of the 2013
  {Conference} of the {North} {American} {Chapter} of the {Association} for
  {Computational} {Linguistics}: {Human} {Language} {Technologies}, Association
  for Computational Linguistics, Atlanta, Georgia, 2013, pp. 847--857.
\newline\urlprefix\url{http://www.aclweb.org/anthology/N13-1105}

\bibitem{coecke_mathematical_2010}
B.~Coecke, M.~Sadrzadeh, S.~Clark,
  \href{http://arxiv.org/abs/1003.4394}{Mathematical {Foundations} for a
  {Compositional} {Distributional} {Model} of {Meaning}}, arXiv:1003.4394 [cs,
  math]ArXiv: 1003.4394.
\newline\urlprefix\url{http://arxiv.org/abs/1003.4394}

\bibitem{zhang_quantum_2018}
P.~Zhang, Z.~Su, L.~Zhang, B.~Wang, D.~Song,
  \href{http://doi.acm.org/10.1145/3269206.3271723}{A {Quantum} {Many}-body
  {Wave} {Function} {Inspired} {Language} {Modeling} {Approach}}, in:
  Proceedings of the 27th {ACM} {International} {Conference} on {Information}
  and {Knowledge} {Management}, {CIKM} '18, ACM, New York, NY, USA, 2018, pp.
  1303--1312.
\newblock \href {http://dx.doi.org/10.1145/3269206.3271723}
  {\path{doi:10.1145/3269206.3271723}}.
\newline\urlprefix\url{http://doi.acm.org/10.1145/3269206.3271723}

\bibitem{arjovsky_unitary_2016}
M.~Arjovsky, A.~Shah, Y.~Bengio,
  \href{http://dl.acm.org/citation.cfm?id=3045390.3045509}{Unitary {Evolution}
  {Recurrent} {Neural} {Networks}}, in: Proceedings of the 33rd {International}
  {Conference} on {International} {Conference} on {Machine} {Learning} -
  {Volume} 48, {ICML}'16, JMLR.org, New York, NY, USA, 2016, pp. 1120--1128.
\newline\urlprefix\url{http://dl.acm.org/citation.cfm?id=3045390.3045509}

\bibitem{wisdom_full-capacity_2016}
S.~Wisdom, T.~Powers, J.~Hershey, J.~Le~Roux, L.~Atlas,
  \href{http://papers.nips.cc/paper/6327-full-capacity-unitary-recurrent-neural-networks.pdf}{Full-{Capacity}
  {Unitary} {Recurrent} {Neural} {Networks}}, in: D.~D. Lee, M.~Sugiyama, U.~V.
  Luxburg, I.~Guyon, R.~Garnett (Eds.), Advances in {Neural} {Information}
  {Processing} {Systems} 29, Curran Associates, Inc., 2016, pp. 4880--4888.
\newline\urlprefix\url{http://papers.nips.cc/paper/6327-full-capacity-unitary-recurrent-neural-networks.pdf}

\bibitem{yuan_speaker_2008}
J.~Yuan, M.~Liberman,
  \href{http://asa.scitation.org/doi/10.1121/1.2935783}{Speaker identification
  on the {SCOTUS} corpus}, The Journal of the Acoustical Society of America
  123~(5) (2008) 3878--3878.
\newblock \href {http://dx.doi.org/10.1121/1.2935783}
  {\path{doi:10.1121/1.2935783}}.
\newline\urlprefix\url{http://asa.scitation.org/doi/10.1121/1.2935783}

\end{thebibliography}
